\documentclass[12pt]{Misc/iopart}

\usepackage{amsfonts,amsbsy,bm}
\usepackage[makeroom]{cancel}

\usepackage{hyperref}
\usepackage{bbm}
\usepackage{array}
\usepackage{bm}
\usepackage{amssymb}
\usepackage{graphicx}
\expandafter\let\csname equation*\endcsname\relax
\expandafter\let\csname endequation*\endcsname\relax
\usepackage{amsmath}
\usepackage{float}
\usepackage{bm}
\usepackage{breqn}
\usepackage{float}
\usepackage{algorithmic}
\usepackage[disable]{todonotes}
\newenvironment{experiment}[1][htb]
  {
   \begin{algorithm}[#1]%
  }{\end{algorithm}}

\usepackage{subcaption}
\usepackage[ruled,vlined]{algorithm2e}

\usepackage{amsthm}

\theoremstyle{definition}

\DeclareMathOperator*{\argmin}{arg\,min}

\begin{document}

\title[Bayesian inference under model misspecification using TL distances]{Bayesian inference under model misspecification using transport-Lagrangian distances:  an application to seismic inversion}

\author{Andrea Scarinci}
\address{Department of Aeronautics and Astronautics, Massachusetts Institute of Technology}
\ead{scarinci@mit.edu}
\author{Michael Fehler}
\address{Department of Earth, Atmospheric,
and Planetary Sciences, Massachusetts Institute of Technology}
\ead{fehler@mit.edu}
\author{Youssef Marzouk}
\address{Department of Aeronautics and Astronautics, Massachusetts Institute of Technology}
\ead{ymarz@mit.edu}
\vspace{10pt}

\begin{abstract}
Model misspecification constitutes a major obstacle to reliable inference  in many inverse problems. Inverse problems in seismology, for example, are particularly affected by misspecification of wave propagation velocities. In this paper, we focus on a specific seismic inverse problem---full-waveform moment tensor inversion---and develop a Bayesian framework that seeks robustness to velocity misspecification. A novel element of our framework is the use of transport-Lagrangian (TL) distances between observed and model predicted waveforms to specify a loss function, and the use of this loss to define a generalized belief update via a Gibbs posterior. The TL distance naturally disregards certain features of the data that are more sensitive to model misspecification, and therefore produces less biased or dispersed \todo{earlier version said ``more accurate"; not sure of best word here} posterior distributions in this setting. To make the latter notion precise, we use several diagnostics to assess the quality of inference and uncertainty quantification, i.e., continuous rank probability scores and rank histograms. We interpret these diagnostics in the Bayesian setting and compare the results to those obtained using more typical Gaussian noise models and squared-error loss, under various scenarios of misspecification. Finally, we discuss potential generalizability of the proposed framework to a broader class of inverse problems affected by model misspecification. 
\end{abstract}

\section{Introduction}
Model error or misspecification is a determining factor in the quality of the solution of an inverse problem. In the context of seismic waveform inversion, for example, this is particularly relevant when it comes to modeling the propagation velocity $\mathbf{V}$ of a seismic wave. Due to the extreme difficulties in characterizing the subsurface medium (e.g., different rock types, three-dimensional spatial inhomogeneities) any velocity model is generally approximate and inaccurate. Mischaracterization of $\mathbf{V}$, however, can impact one's ability to infer other quantities of interest such as the hypo-center $\mathbf{x}$ and the moment tensor $\mathbf{m}$  (focal mechanism) of a seismic event. In deterministic full-waveform-inversion this often results in the well known phenomenon of cycle-skipping, which traps optimizers in local minima \cite{gauthier1986two}. In the Bayesian setting, model misspecification can lead, in a worst case scenario,  to overconfidence in the posterior distribution, i.e., under-reporting of uncertainty \cite{gu2017waveform,kleijn2012bernstein}.

The most direct approach to mitigating the impact of model misspecification is to introduce better physical models (when feasible) or improved statistical discrepancy models. These approaches, however, typically increase computational cost and may compromise parameter identifiability. 

In this paper we investigate instead the benefits of using an alternative, optimal-transport (OT) based, misfit function to measure discrepancies between observed and model predicted data. \todo{First note the use of OT misfits, and go through those examples. Then present TL as a specific case with special properties?} Recent literature has demonstrated the applicability of  OT-based distances to seismic imaging problems in a deterministic setting \cite{engquist2013application,Engquist2017,metivier2018graph,metivier2019graph,yang2018application}. In this context, OT has been shown to produce  drastic reductions in the non-convexity of the objective function, especially when compared to $\ell_p$ distances. A more convex misfit function also implies a more robust solution to the inverse problem when subject to uncertainties in the input parameters. Rigorous mathematical treatment \cite{Engquist2020OptimalTB} has in fact shown that 1-D quadratic Wasserstein distances (a subset of OT distances) are convex functions with respect to dilation and translation when applied to probability density functions. In order for this to remain valid for generic signals as well, it is however necessary to normalize and positivize them accordingly. \todo{Wasserstein formulated how? How specific is that result? Is it for the 1-D problems with positivization via shifting? 1-D W2 with signals positivized and normalized in a slightly different way than the usual }

This last requirement introduces data transformations that are not typically justifiable within the physics of the problem. We therefore decided to focus on a particular case of Wasserstein distance, which does not require signal positivation and normalization and therefore makes it more suitable to deal with seismic waves. Said distance is referred to as the transport-Lagrangian (TL) distance and can be interpreted as the result of solving an optimal-transport (OT) problem between the graphs of two functions. 

While the benefits of using this kind of distance have been already explored in a number of deterministic  inverse problems and applications \cite{thorpe2017transportation}, in this paper we discuss its integration within a fully Bayesian framework. 
\todo{Use a transition that better captures a novelty here: can we say that previous work has not considered statistical, and in particular Bayesian, inverse problem formulations with the TL distance?}
 Within this set-up, we interpret the TL distance as a tool to tackle a broader issue than the cycle-skipping one delineated in the deterministic literature on full waveform inversion (FWI). More precisely, we look at the TL distance as a data ``feature'' extractor that deliberately disregards non-relevant information to the inference of a particular quantity of interest,  minimizing the impact of uncertainties in the model.   

The first step in the development of a coherent Bayesian procedure is to establish a statistical model for the phenomenon of interest and then derive the associated likelihood function. We maintain the classical additive Gaussian noise set-up according to which: \todo{Here and later, we need to deal with $t$ consistently: are $u$ and $y$ functions of $t$, or shall we just discretize time and thus drop the $t$ argument? We need to avoid writing pdfs of a function, as in $p(y(t))$, as this is not mathematically well defined and would require lots of technicality to work around!}
\begin{align}
    \mathbf{y} &= \mathbf{u}(\theta,\mathbf{t}) + \mathbf{e}, \qquad \mathbf{e} \sim \mathcal{N}(0,\Sigma);  \label{eqn:Gaussian_noise_additive_sect1} 
\end{align}
where $\mathbf{y}$ and $\mathbf{u}$ are the vectors containing the discretized versions of the observed  ($y(t)$) and model predicted ($u(t)$) signals respectively, while $\theta$ are the parameters of interest (usually $\theta \in \mathbb{R}^D$). These assumptions naturally lead to the largely used Gaussian likelihood function:
\begin{equation}
    \mathbf{y} \vert \theta \sim \mathcal{N}(\mathbf{u}(\theta),\Sigma). \label{eqn:Gauss_like_sec1}
\end{equation}
This likelihood, however, implicitly assumes the use of the sufficient statistic $T(\mathbf{y,u})=\mathbf{y-u}$ as a summary statistic for data-model discrepancy, which in turns corresponds to an $\ell_2$ norm misfit in the exponent of the Gaussian probability density function (PDF). Instead, while maintaining the noise model outlined in (\ref{eqn:Gaussian_noise_additive_sect1}), we want to introduce the TL distance as our misfit statistic. For such statistic it is not equally straightforward to characterize the conditional distribution of the TL distance given a particular value of $\theta$ and the additive Gaussian noise model in \ref{eqn:Gaussian_noise_additive_sect1}.
 We discuss some results on this topic from recent literature \cite{alquier2016properties}  \cite{bissiri2016general} \cite{motamed2018wasserstein} \cite{dunlop2020stability}. In particular we  discuss the use of the so called Gibbs posteriors and their interpretation in the misspecified context.\todo{I wonder if might be more direct to re-order the arguments in this paragraph, as follows: (i) typically people use additive Gaussian noise models; this leads to the usual conditionally Gaussian likelihood; (ii) instead we want to use TL, but the distribution of the TL distance is not obvious; (iii) so instead we will use the notion of a Gibbs posterior. --> I am not sure I understand exactly where you want the change. Isn't this the current order already? (i made some small changes in any case, idk if they suffice?)}

 We then present an empirical study for the newly derived framework. We focus on the previously described task of moment tensor estimation when the data generating process relies on a different velocity model $\mathbf{V}$ than the one used for modeling predicted waveforms. We show how the TL-based likelihood is  less sensitive to the nuisance effect introduced by the misspecified $\mathbf{V}$ and generally allows for the construction of  more informative posterior distributions on $\mathbf{m}$. We test our framework under different instantiations of velocity misspecification and data-generating moment tensor values.  
 
 Another key part of our study is an assessment of posterior quality under misspecification. In particular we propose a number of quantitative metrics
 to characterize the kind and level of improvement introduced by the TL-based framework. We compare the resulting  posterior distributions to those obtained by using classic $\ell_2$-based Bayesian frameworks such as the one defined in (\ref{eqn:Gaussian_noise_additive_sect1}) and (\ref{eqn:Gauss_like_sec1}). There is no unique way to establish in what respect a posterior distribution is better than another, and if so, how does this depend on the specific use that that the analyst intends to make of it.  
Among the various scoring rules that exist in the literature, we focus on continuous rank probability scores (CRPS) \cite{gneiting2007strictly,gneiting2007probabilistic} as they effectively capture two important qualities a posterior distribution needs to have in order to be used as a practical forecaster: be sufficiently localized (i.e. low variance) and contain the true value of the quantity of interest within its support, preferably in high-probability regions (i.e. low bias). The perfect forecaster would therefore be a delta function that sits right at the true value of the quantity of interest: $\delta(\theta)_{\theta=\theta_{true}}$.      Aside from CRPS, we also discuss ways of checking the self consistency of the inference procedure, specifically in the case of the TL-based likelihood function. For this purpose we focus on recent literature proposed around the concept of rank histograms \cite{talts2018validating} \cite{cook2006validation}. These are checks on the frequentist behavior of Bayesian credible intervals and entirely within-the-model assessment tools. The objective is to verify whether the inference procedure shows any bias in reporting uncertainty around certain regions of the parameter space.         
The rest of this paper is structured as follows:
\begin{itemize}
    \item section 2 will contain a survey of the literature around model misspecification. Optimal-transport-based misfit measures and their use in a deterministic and Bayesian setting will be discussed as well;
    \item section 3 will present the details of Gibbs posteriors as a pseudo-Bayesian update when using the the TL-distance as a misfit measure;
    \item section 4 will contain an application of the defined framework to moment tensor inversion;
    \item section 5 will present and discuss some tools to asses the quality of posterior distributions and apply them to the problem described in section 4;
\end{itemize}
The last part of the paper will summarize the elements of the proposed inference framework and discuss potential generalizations of it to a broader class of inverse problems.

\section{Background and motivation}
%
%
\subsection{Model misspecifictaion in Bayesian inverse problems} \label{sec:LitRewModelMis}
In this section we will briefly recall what is meant by model misspecification and discuss some of the most common approaches to make Bayesian inference more robust. 
The usual approach to Bayesian inference is to assume that the distribution of the data belongs to the family of parameterized distributions defined by the model. More formally, if ${y}_{1:n}$ is a sequence of i.i.d. random variables with density $g(y_{1:n})$ (data distribution, generally unknown) and $f(y_{1:n}|\theta)$ is a family of parameterized densities to approximate $g(y_{1:n})$, with $\theta \subset \Theta$, we say the model is well-specified if there exists a $\theta_0$ such that  $g(y_{1:n}) = f(y_{1:n} | \theta_0)$. Under such premises (and other, minor conditions), the Bernstein–von Mises theorem holds \cite{kleijn2012bernstein}, which ensures that the posterior distribution, with an infinite amount of data, becomes asymptotically normal and centered around the true parameter value $\theta_0$. Additionally, the posterior credible intervals are ensured to have good frequentist coverage. On the contrary, when the model is misspecified i.e., $g(y_{1:n}) \neq f(y_{1:n} | \theta)$ for any choice of $\theta$, the posterior distribution will become asymptotically normal centered around a value $\theta^*$ which is:
\begin{align}
	\theta^* = \arg \min_\theta \text{ } \mathcal{D}_{KL}  \left (g(\cdot)  \, ||  \, f (\cdot \vert \theta) \right ).
\end{align}
where $\mathcal{D}_{KL}$ is the Kullback-Leibler (KL) divergence. Being the minimizer of the KL divergence does not ensure that the model distribution will be able to reproduce the data. Moreover, the KL divergence does not necessarily present a unique minimum.  The variance will still shrink as far as new data is incorporated and the posterior distribution may therefore provide a misleading characterization of the uncertainty implicit in the problem. \\

A certain amount of research has been conducted on how to make inference more robust to model misspecification \cite{white1982maximum,hu1998fit,chatfield1995model}. The classical approaches can be categorized in two threads \cite{smith2013uncertainty}: 1) better physical modelling: either in terms of the actual physics of the phenomenon  or in terms of model selection/extension \cite{ray2017low,ray2016frequency,buckland1997model,schorfheide2000loss}; 2) better/more robust statistical modeling by extending the additive Gaussian noise model with a term $\delta(\phi,t)$ that aims at modelling additional statistical relations between the observed data and model ($\phi$ being additional parameters, not necessarily related to $\theta$) \cite{kennedy2001bayesian}. Both strategies inevitably increase the computational cost and are prone to suffer from parameter identifiability issues. In addition, by calibrating $\theta_{dis}$ through the data, it may be difficult to discern whether the term $\delta$ is only compensating for missing statistical modeling or it is instead acting as a compensator for the physical model itself, which would result in classic overfitting issues.  \\

More recent approaches have departed from these classical trends. The concept of  \textit{coarsening}  \cite{miller2018robust} for example, modifies the standard Bayesian approach to introduce a  posterior distribution obtained by conditioning not on the event that the data is generated by the model distribution, but instead on some measure of discrepancy between the observed data and model predicted. This discrepancy measure is often taken to be the Kullback-Leibler (KL) divergence between the two (empirical) distributions of the data. Intuitively, this construct produces the following effect: as long as some discrepancy is observed between the observed and model predicted data, the posterior distribution will not  concentrate around a specific value, even with an infinite amount of data. This property avoids the undesired behavior described by the Bernstein-Von-Mises theorem under model misspecification \cite{kleijn2012bernstein}. \\

Another line of research on model misspecification has its roots in decision theory \cite{muller2013risk,watson2016approximate,grunwald2018safe}. Here, robustness to model misspecification is assessed by means of a minimax rule. In \cite{watson2016approximate} a loss function $L(\theta)$ is defined with the model parameter $\theta$ as an argument. The robustness of the model to misspecification is then assessed by calculating how far, in a KL sense, can the true posterior be from the estimated one, while maintaining some upper bound on the loss function.  The definition of the loss function itself and the maximum acceptable radius in which the perturbed posterior can lay,  have to be tied to the features the analyst cares about. \\
Also based on decision making type strategies is the recent concept of ``safe Bayes'' \cite{grunwald2018safe}. In this work the central idea is that a posterior distribution $p(\theta \vert y_{1:n})$ can  be considered ``reliable'' in  estimating a certain quantity of interest $\theta$ if $\mathbb{E}_{p(\theta | \mathbf{y})}[\theta]$ is equal to the same expectation taken with respect to another distributions $p$ belonging to a broader ``credible'' set of distributions. This set is supposed to be built so that it subjectively contains the true distribution for $\theta$. The concept of robustifying the inference process by looking  at summary statistics of the quantity of interest is somewhat relatable to the objective discussed at the beginning of this paper: to build an appropriate misfit function that only takes into account features of the data relative to $\theta$. \\

In this paper, by focusing on a seismic inverse problem, we take a different perspective  than the one offered by the methodologies discussed so far. Given that the model complexity of seismic wave propagation is already high enough, more sophisticated modeling will not be the path to follow. Coarsening  does make inference more robust to model misspecification while keeping the complexity of the model fixed. However, it does so in a generalized fashion,  by increasing the variance of the posterior distribution  without taking into account whether, for at least a subset of the quantities of interest, it is still possible to capture the true parameter value. Decision theoretic frameworks, although theoretically appealing, require a number of assumptions to define with respect to what the posterior shall be considered robust to.  How to build credible sets and loss functions can in fact be a challenging problem by itself and the result consists again in adding more uncertainty to the posterior distribution, rather than directly tackling  model misspecification.

\subsection{Optimal transport as misfit measure}
\todo{Again some shortening here. I think the plan here should be to mention past work that uses OT (of various kinds) to build a misfit function in estimation/inference/inverse problems and then get right into the TL distance specifically. It's only on this last point that we need to define things precisely and with all the equations.\\
Also, one thing that strikes me here is that we are already in a setting where we have assumed that the data are "signals" or waveforms, i.e., functions of time. If that is indeed the scope, then let's make it explicit!}

Central to any inverse problem, both in the deterministic and Bayesian framework, is the choice of a misfit function to compare model and predicted data. We have described in the introduction how choosing this metric can play a determining role in performing good inference, especially when dealing with time series. The following discussion  will therefore focus on this specific data-type, except when stated otherwise. \\

The most recurrent choice is the $\ell_2^2$ norm (least-squares), which is also implicitly obtained by adopting the traditional model that defines observations as model prediction plus Gaussian noise. Being an $\ell_{p}$-norm, the Euclidean distance compares two data vectors element-wise. This represents a limitation when data points are signals (even though discretized), since they inherently exhibit a temporal-spatial structure that cannot be captured by simply comparing them along the time domain. In fact $\ell_p$ norms  ignore the correlation that exists between different points of the signals and can provide rather distorted distance measures.

Some literature exists about possible choices for distance measures between time signals  for a variety of purposes such as pattern recognition, signal classification, detection etc \cite{lauwers2017time}. Some still make  use of the $\ell_2$-distance, but only on slices of the signals, or after performing a circular shift of the time domain \cite{gharghabi2018ultra}. Other approaches propose counting the number of subsequences of the signals that are similar in an $\ell_2$ sense \cite{gharghabi2018ultra}. Parameterizations of the signals (i.e., low-rank approximation) have also been proposed such that the comparison is made in this alternative domain rather in the original time or space one \cite{basseville1989distance, chan2008wavelet}. While attractive, these techniques are only relevant to specific  applications and therefore tend to have limited applicability.

\subsubsection{The optimal transport problem}

Recent advances in the domain of optimal transport and its many applications have lead a number of contributions in the field of signal analysis. Optimal transport allows the type of across-coordinate comparison of functional data being sought, with some distinctive features compared to the more commonly known dynamic time warping. Classical optimal transport (OT) is defined between two probability measures (Kantorovic problem) and solves an assignment problem between the masses at each point of the distributions. Often, the cost function is a distance between the points in the support of the distribution, in which case it is referred to as $p$-Wasserstein distance. A distinctive feature of Wasserstein distances vs. dynamic time warping is that causality is not ensured. This may seem a limitation in its application to signal comparison because of the inherent sequential nature of time signals. However, when dealing with model misspecification this aspect can actually be beneficial in that inconsistencies in the modeling can produce anticipation or delay in the reproduction of some parts of the observed signal.   \\

 For the OT problem to even have a solution it is however necessary for the input signals  to be normalized and positivized. These conditions are naturally satisfied by probability measures, signals however are not measures i.e., they do not sum up to 1 and are not non-negative. A common workaround to this problem is to shift the signal along the ordinate axis to make it positive and then divide it by the sum of all of its points  \cite{Engquist2017,Viriuex2016,thorpe2017transportation}. Having the signal transformed in such a way also allows a fast, analytical, computation of the Wasserstein distance in 1-D.   
Successful attempts of using the Wasserstein distance in this fashion have been made in the field of waveform inversion too \cite{Engquist2017,Viriuex2016}. Promising results were achieved in these works for velocity inversion. As mentioned in the introduction, the use of OT-based misfit functions has proved to be beneficial in terms on non-convexity reduction and cycle-skipping effects. \cite{brossier2015velocity,warner2013full}. \\

While computationally convenient, the transformation of the signals that is required appears somewhat artificial and not justified by the physics of the problem. In addition the transformation can distort the signal, smoothing out amplitude vs frequency differences \cite{thorpe2017transportation}.  One approach that avoids this kind of pitfalls is to use the so called transport-Lagrangian (TL) distance, which is a specific instantiation of the Wasserstein distance with the following formulation:
\begin{align}
    TL^{\lambda}_p(\mathbf{a(x)} || \mathbf{b(y)}) &= \min_{P_{i,j}} \sum_{i,j} C_{i,j} P_{i,j}; \nonumber \\
    \text{s.t.} \,\,\,\,\, \sum_{j=1}^M P_{i,j} &= \frac{1}{N}; \nonumber \\
     \,\,\,\,\,\,\,\,\,\,\,\,\, \sum_{i=1}^N P_{i,j} &= \frac{1}{M}; \label{eqn:TL-dist}\\
     P_{i,j} &\geq 0; \,\, P \in \mathbb{R}^{N \times M}; \nonumber \\
     C^{\lambda}_{i,j} &=  \lambda * |x_i - y_j|^p + |a(x_i) - b(y_j)|^p; \,\, C \in \mathbb{R}^{N \times M}. \nonumber 
\end{align}
where: $\mathbf{a} \in \mathbb{R}^N$ and $\mathbf{b} \in \mathbb{R}^M$ are two vectors representing the discretized version of two functional data-points over the respective coordinates $\mathbf{x,y}$; $C$ is a cost matrix; $P$ a transport plan matrix and $\lambda$ a weighing parameter between horizontal and vertical cost.
This formulation can be interpreted in two different ways:
\begin{enumerate}
    \item optimal transport between two 2-D uniform measures defined on $\{x_i \times a(x_i)\}$ and $\{ y_j \times b(y_j)\}$ i.e., OT between Lebesgue measures raised onto the graphs of $a$ and $b$ $\tilde{a} (x_i \times a(x_i)) = \mathcal{L}(x_i \times a(x_i))$ and $\tilde{b} (y_j \times b(y_j)) = \mathcal{L}(y_j \times b(y_j))$ ;
    \item optimal transport between two (1-D) uniform measures where the cost is defined as $C^{\lambda}(x_i,y_j) = \lambda |x_i - y_j|^p +  |a(x_i) - b(y_j)|^p$;   
\end{enumerate}

This distance is particularly interesting as it avoids unnatural data transformations while still allowing an OT formulation.  In addition, while computing the Wasserstein distances amounts to solving a linear programming (LP) problem ($O(N^3)$ complexity, with $N$ being the dimension of the data-vector), for the special case of the TL distance one can adopt more specialized algorithms that solve an assignment problem (i.e., $N=M$, and $P$ are permutation matrices). The algorithm of our choice is the auction algorithm \cite{bertsekas1981new}, which exhibits a nearly quadratic complexity or average complexity of $O(N^2 log(N))$  for problems with $N < 1000$ \cite{schwartz1994computational,metivier2019graph} .\\

Rigorous discussion on the applicability of the TL measure as an objective function in deterministic inversion has been conducted in \cite{metivier2019graph}.  Improvements in the convexity of the misfit have emerged as the primary effect of the choice of such a distance measure. \\

In this paper we intend to define a strategy for the use of the TL in a Bayesian setting.

\section{Transport-Lagrangian based likelihood function}
\todo{This section repeats things from the intro a bit too much; take a pass to eliminate redundancy.}
In this section we intend to answer the following question: how can we build a coherent Bayesian framework  around the  TL distance as a misfit statistic? While maintaining the classic set-up of additive Gaussian noise, we seek an alternative expression for the likelihood  $p(\text{TL}^2(\mathbf{y,u})\vert \theta)$ (where $\mathbf{y}$ and $\mathbf{u}$ are the vectors containing the discretized form of the observed and model predicted signals, while $\theta$ are the model parameters).  Also note the choice of $p=2$  to allow for more direct comparison with the $\ell_2$ norm as a misfit statistic. \\

At this point, there are two main impediments that stand in the  way of defining a coherent Bayesian framework for a TL-misfit, both in a well-specified and misspecified setting. First, calculating the TL-distance involves solving an optimal transport problem, which implies, in turn, a minimization problem: this non-linearity makes it difficult to derive an analytical expression for the likelihood  $p(\text{TL}^2(\mathbf{y,u})\vert \theta)$. In the second place, we stated multiple times that it is of our interest to evaluate the robustness of such misfit measure in a misspecified context. From a rigorous standpoint however, the definition of likelihood assumes a context in which the data come exactly from the specified model. Therefore, even if we were able to obtain an exact expression for $p(\text{TL}^2(\mathbf{y,u})\vert \theta)$, this would not mean that the same expression could be used in a misspecified context without introducing some sort of inconsistency. \\

We therefore propose to adopt an alternative framework in which both misspecification and the newly introduced misfit measure can be integrated. In the statistical literature, the posterior distribution obtained through this framework is typically referred to  as the Gibbs posterior. \todo{For the purpose of this paper, I don't think we should so much cast the Gibbs posterior as a second poorer option! I think we should just say that we use it!}
A full derivation is contained in \cite{bissiri2016general}, but we recall here a short summary. The central idea is to define a loss function $\mathcal{L}(\pi,\mathbf{y};p)$ over our prior beliefs $\pi(\theta)$,  observations $\mathbf{y}$ and space of probability measures $p$ over $\theta$. We then claim that a valid update of our beliefs based on available data is given by:
\begin{align}
    \hat{p}&=\argmin_{p}\mathcal{L}(\pi,\mathbf{y};p) .
\end{align}
This claim is justified by the argument that, in general, between two measures $p_1$ and $p_2$, one would naturally prefer the one that produces a lower value of the loss function, given the same data-set. The authors also choose a specific expression for the loss function that contains both of the fundamental ingredients of a Bayesian update i.e. balance between prior information (Kullback-Leibler divergence between $p$ and $\pi(\theta)$) and adherence to observed data under the form of expected loss:
\begin{align}
    \hat{p}&=\argmin_{p} \left ( \int \ell(\theta,\mathbf{y}) p(d\theta) + \mathcal{D}_{KL}(p,\pi(\theta)) \right ).
\end{align}
The function $\ell(\theta,\mathbf{y})$ is a generic measure of model-data discrepancy (more discussion later). The authors show that the minimizer $\hat{p}$ takes the form:
\begin{equation}
    \hat{p}(\theta ) = \frac{\exp\{-\ell(\theta,\mathbf{y})\} \pi(\theta)}{\int \exp\{-\ell(\theta,\mathbf{y})\} \pi (d\theta)}. \label{eqn:belief}
\end{equation}
This expression can justify a prior-to-posterior update through an exponential form given a generic loss function (or misfit measure) $\ell(\theta,\mathbf{y})$. While not a rigorously Bayesian update, it still captures the two main ingredients of Bayesian inference and provides a rigorous argument for using an exponential pseudo-likelihood. Additionally,  we note that if it is known that the  data arose from a given family of distributions  (e.g. $p(\mathbf{y}\vert \theta)$), then equation (\ref{eqn:belief})  reverts exactly to Bayes formula, by taking $\ell(\theta,\mathbf{y})=-\log(p(\mathbf{y}\vert \theta))$. This ensures the expression above constitutes a rational update with any misfit measure  both in the well and misspecified context. 

In our experiments, we adopted a specific expression for the Gibbs posterior as outlined in  \cite{motamed2018wasserstein} (already experimented in a seismic inverse problem in \cite{scarinci2019robust}): \todo{Shouldn't this just be $\ell(\theta, \mathbf{y})$ below? --> I think not, because the $\ell$ is intended to be just a distance measure, while the data-update is the "alternative" posterior. They do the same in the Bissiri paper unless I got something wrong   }
\begin{equation}
    p(\mathbf{y}|\theta) = s^N  \exp \left (-s \, \text{TL}_p\left(\mathbf{y},\mathbf{u}(\theta) \right ) \right ) \label{eqn:likelihood_EXP_TL}.
\end{equation}
where, $N$ is the number of observations while the parameter $s$ acts as scaling factor.
\paragraph{The role of the $s$ parameter} This parameter plays no role in the data-generating process and has no physical meaning,\todo{Maybe the scale itself does have physical meaning? In general i think we had come to the conclusion it does not since it whatever we insert as the argument of the exponential. In any case it is not just noise.} but it is necessary to ensure the values taken by the loss functions (in this case the TL-distance) are of the right order of magnitude to produce meaningful posterior distributions after being exponentiated.  The scaling is therefore not an ad-hoc manipulation of the data to achieve more desirable results, bur rather a necessary adjustment to integrate any given loss function with a prior-to-posterior update that is not derived explicitly from a physical model. This is reflected in the computational scheme used to calibrate the amount of scaling: $s$ can be treated as a hyper-parameter and estimated through a hierarchical Bayesian framework. We associate to $s$ a Gamma distribution as a conjugate prior, which allows a Gibbs update \cite{motamed2018wasserstein} in a Markov chain Monte Carlo (MCMC) algorithm that otherwise uses generic adaptive Metropolis \cite{haario2001adaptive}  for $\theta$ updates. The choice of values for the shape and rate parameters of the Gamma prior is particularly critical to obtainment of a meaningful posterior. These values need to be picked in such a way that whatever loss function $\ell(\theta, \mathbf{y})$ is chosen to be used in the Gibbs posterior, it will scaled appropriately to avoid $\exp{(-s \cdot \ell(\theta, \mathbf{y}))}$ being numerically insensitive to different values of $\theta$, making inference unfruitful. In the following section we will discuss the reasoning behind the choice of the Gamma prior for $s$ through a  numerical example.

\section{Application: seismic moment tensor inversion}
In this section we present an application of the TL distance to a seismic inverse problem called moment tensor inversion. We will dedicate the first sub-section to the description of the forward model and follow with a specific instantiation of the inverse problem. A more systematic investigation of the effect of the TL-distance in a misspecified setting will be conducted in section \ref{sec:CRPS and Histpograms}.

\subsection{The seismic inverse problem}\label{intro:problem}
 
A major goal in seismology is to understand how seismic waves propagate through a given terrain (\textit{forward-problem}). Parallel to this is the so called \textit{seismic inverse problem}, which relates the observed seismic displacements (typically recorded by seismograms on the Earth's surface) to their source (earthquake). Characterizing earthquakes provides better understanding of the earth structure and is of particular interest in the oil and gas as well as geothermal industries, where small earthquakes are artificially provoked  to detect the presence of a reservoir in a terrain of interest. The equation that is at the heart of this problem is the momentum equation for an elastic continuum:
\begin{equation}
	\rho \frac{\partial^2 u_i}{\partial t^2} = \sum_{j=1}^{3} \frac{\partial \tau_{ij}}{\partial x_j} + f_i, \label{eqn:elasticWave}
\end{equation}
where: $\rho =$ medium density; $u_i =$ displacement in the direction $i$, $\tau_{ij} = ij$-th element of the stress tensor, $f_i =$ body force along direction $i$.  
	
Solving (\ref{eqn:elasticWave}) for the displacement $u_i$ is generally difficult \cite{shearer2009introduction}. One approach to this PDE is to express the solution $u_i(\mathbf{x_s},$ (displacement at location $\mathbf{x_s}$) as:
\begin{equation}
	u_i(\mathbf{x_s},\mathbf{t}) = \mathbf{G_i}(\mathbf{x_s},\mathbf{t}) \cdot \mathbf{m}^T, \label{eqn:elasticWaveSolution1}
\end{equation}
where $\mathbf{G_i}$ is the Green's function, i.e., the solution of the PDE when a unit force is applied at $\mathbf{x_s}$ (earthquake or source location), and  $\mathbf{m}$ is the \textit{moment tensor}, which represents the force couples that can provoke an earthquake. In its most general form it is a $3 \times 3$ symmetric matrix, meaning only 6 of its elements are independent. This allows to recast it as a $1 \times 6$ vector $\mathbf{m}$ as indicated in (\ref{eqn:elasticWaveSolution1}), where $\mathbf{G_i}$ is also a $1 \times 6$ vector. Further simplifications and decompositions are possible when the earthquake mechanisms are restricted to be of a particular type (double couple). \\

The Green's function $\mathbf{G_i}$ contains, implicitly, all the information relevant to the seismic phenomenon, beyond the source term. This includes quantities such as the location of the earthquake $\mathbf{x}$, the propagation velocity field of the compressional and shear waves ($V_p(\mathbf{x})$ and $V_s(\mathbf{x})$), the medium density $\rho$, etc. The objective of full seismic waveform inversion is that of inferring one or a subset of these quantities of interest, given the observed displacements $y_i(\mathbf{x_s},\mathbf{t})$ (normally recorded through seismograms positioned at given locations on the field of interest). A typical choice is to invert for the velocity model. Even though the velocity is in general not homogeneous with respect to $\mathbf{x}$,  in most applications it is restricted to assume some fixed values within a certain portion or layer of the terrain of interest, reducing the complexity of the model. For this reason the notation is simplified to $V(\mathbf{x}) = \mathbf{V}$. \\

For our application, however, we will focus on estimating the moment tensor components, while considering all other parameters (particualrly $\mathbf{V}$ and $\mathbf{x}$) known and fixed to a given value. We will therefore generally refer to the Green's function by making explicit its dependence to the location of the source earthquake and the velocity model only: 
\begin{equation}
	\mathbf{u}(\mathbf{t}) = \mathbf{G}(\mathbf{x},\mathbf{V},\mathbf{t}) \cdot \mathbf{m}^T, \label{eqn:elasticWaveSolution}
\end{equation}
where $\mathbf{u}(\mathbf{t})$ is the vector containing the displacements as a function of time, along all directions and at all locations of interest.  The Green's function $\mathbf{G}$ is a non linear operator in $\mathbf{x}$ and $\mathbf{V}$. This implies that the objective function of a typical least squares minimization problem will most likely be non-convex and that, in a Bayesian setting, the full posterior distribution $p(\mathbf{X,V,m}\vert \mathbf{y})$ will be non-Gaussian.

\subsection{General experiment set-up}
 We are interested in evaluating the benefits of using the TL distance as a misfit statistic when solving the moment-tensor inverse problem in presence of model misspecification. 
To this purpose, we conduct an experiment using synthetically generated data from a layered-media model. This model assumes that the waves travel through homogeneous elastic layers of different depth and velocity (one value for the velocity  $V_p$ of the primary waves and one value $V_s$ for the velocity of the secondary waves). The adopted solution to the forward problem is described in \cite{bouchon1977discrete}. In this application we assume to collect data from 4 different stations in each displacement direction (N-E-Z). These stations are located at $z=0\text{\, m}$ from the Earth surface and we can express their positions with respect to the epicenter of the earthquake in polar coordinates: station 1 - $r_1=5.6 \text{\, km}, \theta_1 = 60 \deg$; station 2 - $r_2=3.5 \text{\, km}, \theta_2 = 110 \deg$; station 3 - $r_3=4.1 \text{\, km}, \theta_3 = 250 \deg$; station 4 - $r_4=5.3 \text{\, km}, \theta_4 = 280 \deg$. The source is located at 1.1 km depth with duration 0.01 s. The entire waveform is recorded for 8.192 s, while we will only use the portion between 1 s and 7 s for the inversion. Model misspecification will be introduced by using a different velocity model for the data-generating process vs. the inference process. \\

We will now describe the velocity models that will be used throughout this paper to create both well-specified and misspecified inference settings. In all of our experiments we will use a four-layered media velocity model (table \ref{tab:4layer_vmodel}) for the inference process. In a realistic setting, this model represents the analyst's best attempt at describing the geophysical characteristics of the terrain of interest. We call this model $\mathbf{V}_{4_{lay}}$. For the data-generating process we will instead use two variations of $\mathbf{V}_{4_{lay}}$ to introduce misspecification.
In particular we will test two modes of misspecification:
\begin{itemize}
    \item {\textbf{mode 1:}} Data is generated  through a model that exhibits 3 layers instead of 4, as specified in table \ref{tab:3layer_vmodel}. We call this model $\mathbf{V}_{3_{lay}}$; 
    \item {\textbf{mode 2:}} The model used for data generation has $20\%$ variations around the velocity values used in $\mathbf{V}_{4_{lay}}$. We call this model $\mathbf{V}_{4-20\%}$.
\end{itemize}

\begin{table}[H]
\centering
\begin{tabular}{||c c c c c c c||}
\hline
Nbr. & Thickness & $V_p $ & $V_s$ & $\rho$  & $Q_p$ & $Q_s$  \\ [0.5ex] 
& \scriptsize $km$ &\scriptsize $km/s$ & \scriptsize $km/s$ & \scriptsize $g/cm^3$& & \\
\hline\hline
\textbf{1} & 0.5 & 2.5 & 1.00 & 2.0 & 40 & 20 \\
\hline
\textbf{2} & 0.5 & 3.0 & 1.50 & 2.0 & 40 & 20 \\ 
\hline
\textbf{3} & 0.5 & 3.5 & 1.75 & 2.0 & 40 & 20 \\
\hline
\textbf{4} & 1.0 & 5.5 & 2.75 & 2.0 & 40 & 20 \\
\hline
\end{tabular}
\caption{Layer model used for data generation. }
\label{tab:4layer_vmodel}
\end{table}

\begin{table}[H]
\centering
\begin{tabular}{||c c c c c c c||}
\hline
Nbr. & Thickness & $V_p $ & $V_s$ & $\rho$  & $Q_p$ & $Q_s$  \\ [0.5ex] 
& \scriptsize $km$ &\scriptsize $km/s$ & \scriptsize $km/s$ & \scriptsize $g/cm^3$& & \\
\hline\hline
\textbf{1} & 0.8 & 2.5 & 1.00 & 2.0 & 40 & 20 \\
\hline
\textbf{2} & 1 & 3.2 & 1.60 & 2.0 & 40 & 20 \\ 
\hline
\textbf{3} & 0.7 & 5.5 & 2.75 & 2.0 & 40 & 20 \\
\hline
\end{tabular}
\caption{Layer model used in inference mode 2}
\label{tab:3layer_vmodel}
\end{table}
The objective is to test whether the TL distance performs better in terms of recovering the correct value of the moment tensor $\mathbf{m}_{true}$ compared to the the implicitly induced $\ell_2$ norm of the additive Gaussian model (as described in (\ref{eqn:Gaussian_noise_additive_sect1}) and (\ref{eqn:Gauss_like_sec1})). 

\subsection{Experiment 1 set-up}
We first test the TL distance by examining its behavior in a mode 1 misspecification setting as described in the experiment \ref{alg:experiment1} prospect.\\
\begin{experiment}[H] \label{alg:experiment1}
\SetAlgoLined
{
 Set $ (\text{Strike}, \text{Dip}, \text{Rake}) = (300^\circ, 20^\circ, 150^\circ) \rightarrow \mathbf{m}_{true} = \text{\scriptsize{ [ -0.50, 0.18, 0.32, 0.01, 0.74, -0.51]}}  $\;
 Generate data $\mathbf{y}$ \footnote{one vector containing the concatenated signals for each station} according to: \\
\vspace{0.1cm}
\hspace{2cm}
$
\mathbf{y} = \mathbf{G}(\mathbf{x}_{true},\mathbf{V}_{3_{lay}},\mathbf{t}) \cdot \mathbf{m}_{true}^T + \mathbf{e} \text{\,\,\, where: } \mathbf{e} \sim \mathcal{N}(0,\sigma^2 \mathbb{I});
$
\vspace{0.2cm}

Estimate the posterior $p(\mathbf{m} \vert \mathbf{y} )$ assuming the following model for the data:\

\vspace{0.3cm}
\hspace{2cm}
$
(\text{{mode 2}) \,\,\,} \mathbf{u} = \mathbf{G}(\mathbf{x}_{true},\mathbf{V}_{4_{lay}},\mathbf{t}) \cdot \mathbf{m}^T + \mathbf{e} \text{\,\,\, where: } \mathbf{e} \sim \mathcal{N}(0,\sigma^2\mathbb{I}); 
$ \\
\vspace{0.2cm}
\hspace{1.0cm} where $\sigma$ is known and set to $\sigma = 10^{-3}$.
\vspace{0.1cm}
}
\caption{Inference procedure}
\end{experiment}

The posterior will be a joint posterior over the $6$ dimensional space of the moment tensor components (generally correlated). In order to evaluate the impact of the choice of the misfit statistic on the solution of the problem just described, we will integrate both the classic  $\ell_2$ misfit measure as well as the TL-based distance into the pseudo-likelihood presented in the previous section. More explicitly:
\begin{align}
    \ell_2: \,\,\,\,\,\, p( \mathbf{y}^k \vert \mathbf{m}^k) &\propto \exp{\left( -s \, \vert \vert \mathbf{y}^k - \mathbf{u}^k \vert \vert_2^2 \right)}; \\
    TL_2: \,\,\,\,\,\, p( \mathbf{y}^k \vert \mathbf{m}^k) &\propto  \exp{\left( -s \text{\, TL}_2(\mathbf{y}^k,\mathbf{u}^k) \right)};
\end{align}
The TL-based likelihood is the one derived in (\ref{eqn:likelihood_EXP_TL}), where the parameter $s$ acts as a scaling parameter. The $\ell_2$-based likelihood is derived by simply substituting the $\ell_2$ to the $\text{TL}_2$ misfit in the Gibbs posterior, although it is important to note that its analytical form corresponds exactly to the one that could be obtained by conditioning on the model parameters, starting form equations (\ref{eqn:Gaussian_noise_additive_sect1})(\ref{eqn:Gauss_like_sec1}). In this case the $s$ parameter could be directly interpreted as the model variance and would not need to be estimated through a hierarchical procedure (if assumed to be known). For consistency with the TL case, however, we treat it as a hyper-parameter and leave the discussion for the analytical solution of the linear Gaussian inverse problem for section \ref{sec:analytical}.
At the end of the experiment we will therefore have one posterior for each of the following cases:
\begin{align}
     \text{mode 1 - $\ell_2$: \qquad}    &p_{\ell_2}(\mathbf{y}^k \vert \mathbf{m}^k);    \\
     \text{mode 1 - $\text{TL}_2$: \qquad } &p_{\text{TL}_2}(\mathbf{y}^k \vert \mathbf{m}^k).    
\end{align}
Before discussing the results  we briefly describe the settings for the actual algorithm used for Bayesian inversion.
\paragraph{Algorithm} As already mentioned before, we implemented a Gibbs-within Metropolis scheme that updates $\mathbf{m}$ with a classic adaptive MCMC step, while for $s$ it exploits the conjugacy of the Gamma prior. In particular we can sample $s$ through a Gibbs update, meaning we can sample from the full conditional $p(s\vert \mathbf{m,y}) = \text{Gamma}(a,b+\ell(\mathbf{y,u}))$. The term $\ell(\mathbf{y,u})$ stands for whichever distance measure we are considering, either $\ell_2$ or $\text{TL}_2$. The coefficients $a$ and $b$ are the shape and rate parameter of the Gamma prior on $s$.
\paragraph{Prior on moment tensor} For the moment tensor prior we adopt a uniform distribution\todo{Is it a bounded uniform distribution as written here, or is it a flat (improper) prior on all of $\mathbb{R}^6$? I forget!} on the 6-dimensional $\ell_{\infty}$ unit ball (i.e. $\mathcal{U}(||\mathbf{m}||_{\infty}\leq 1)$).

\paragraph{Prior on $s$} We set a Gamma prior on the scaling parameter $s$: $\text{Gamma}(a,b)$. The rationale behind the choice for the shape ($a$) and rate ($b$) parameters is as follows: in order for $\exp{\{- s \cdot \ell(\mathbf{y,u}) \}}$ not to concentrate  around $0$ or $+\infty$ for any value of proposed $\mathbf{m}$, the monomial $s \cdot  \ell(\mathbf{y,u}) $ needs to take values within the range $[-10, 10]$, at least for a subset of  $||\mathbf{m}||_{\infty}\leq 1$. Depending on the average magnitude of the distance measure $\ell(\mathbf{y,u})$, the Gamma prior must be chosen such that:
    \begin{equation}
        \mathcal{O}(s_{post}) \cdot \mathcal{O}(\ell(\mathbf{y,u})) \approx \mathcal{O}(1)
    \end{equation}
    where  $s_{post}$ is the $s$ sampled from the conjugate posterior (i.e. $ s_{post} \sim \text{Gamma}(a, b + \ell(\mathbf{y,u}))$);
In our experiments, for  both modes of misspecification we have that:
    \begin{equation}
        \mathcal{O}(\ell(\mathbf{y,u})) \approx 10^{-2}
    \end{equation}
    which in turn requires:
    \begin{equation}
        \mathcal{O}(s_{post}) \approx 10^2
    \end{equation}
 Since $\mathbb{E}(s_{post})= a \cdot \left( b + \ell(\mathbf{y,u})  \right)^{-1} $ 
    and $\mathbb{V}(s_{post}) = a \cdot \left( b + \ell(\mathbf{y,u}) \right)^{-2}$, then an appropriate choice for the shape and scale parameter would be $a=100, b=1$. Given that $\mathcal{O}(\ell(\mathbf{y,u})) \approx 10^{-2}$, this will result in:
    \begin{equation}
        \mathbb{E}(s_{post}) = 10^2
    \end{equation}
    and approximately equal value for the variance. 

\subsection{Results}
We want to compare two sets of 6 posterior distributions $p_{\ell_2},p_{TL_2}$ and understand if and how the TL-based likelihood performed better than the $\ell_2$ based one. At this stage we can afford to visually inspect each of the posterior distributions and provide a qualitative judgement. In the following section however, we will broaden the experimental set-up and discuss ways of carrying out more systematic and quantitative evaluations of the quality of the posterior distributions. \\

\begin{figure}
    \centering
    \includegraphics[width=\textwidth,trim={0cm 0.5cm 0cm 0cm},clip]{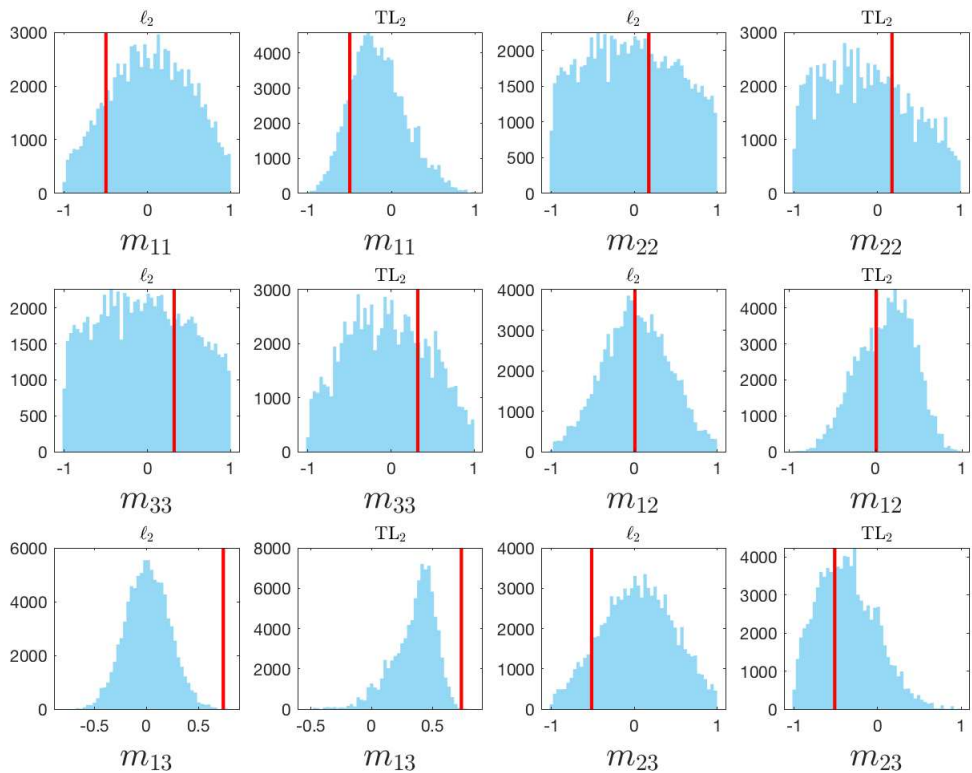}
    \caption{Sample $p_{\ell_2}, p_{TL_2} $ posteriors for mode 1 misspecification }
    \label{fig:sample_posteriror_mode1}
\end{figure}

In Figure \ref{fig:sample_posteriror_mode1} the marginals of $p_{\ell_2},p_{TL_2} $ for each moment tensor component are shown side-by-side to facilitate comparison. It is quite clear how the TL-based posteriors seem to provide a better representation of the uncertainty around the true parameter values (red-lines). By ``better'' we mean in this case that TL-based posteriors are usually more centered around the truth and exhibit less spread around it. In contrast the $\ell_2$-based posteriors are almost uniform for some parameters ($m_{33},m_{22}$) or completely off-centered for others ($m_{13}, m_{23}$).  
{}

These figures however represent one specific instantiation of the problem and are therefore  only anecdotal. In the following section a more systematic investigation of the behavior of the TL-distance compared to the $\ell_2$ misfit will be conducted.  In particular we will attempt to  answer the more fundamental question of how to evaluate the quality of posterior distributions and, more concretely, how to compare or rank them. 

\section{Assessing and comparing posterior quality} \label{sec:CRPS and Histpograms}

While Bayesian inference has become a widely used  in many applications, it is still not entirely clear what constitutes a ``good'' posterior: how much uncertainty is the right amount of uncertainty? Should the true value of the parameter always be expected to lie in high probability regions of the posterior (e.g. at the center of a Gaussian posterior)? A number of answers exists  in literature and their content largely depends on the more fundamental question: ``What do we want to use the posterior for?''. \\

\begin{figure}
    \centering
    \includegraphics[width=\textwidth]{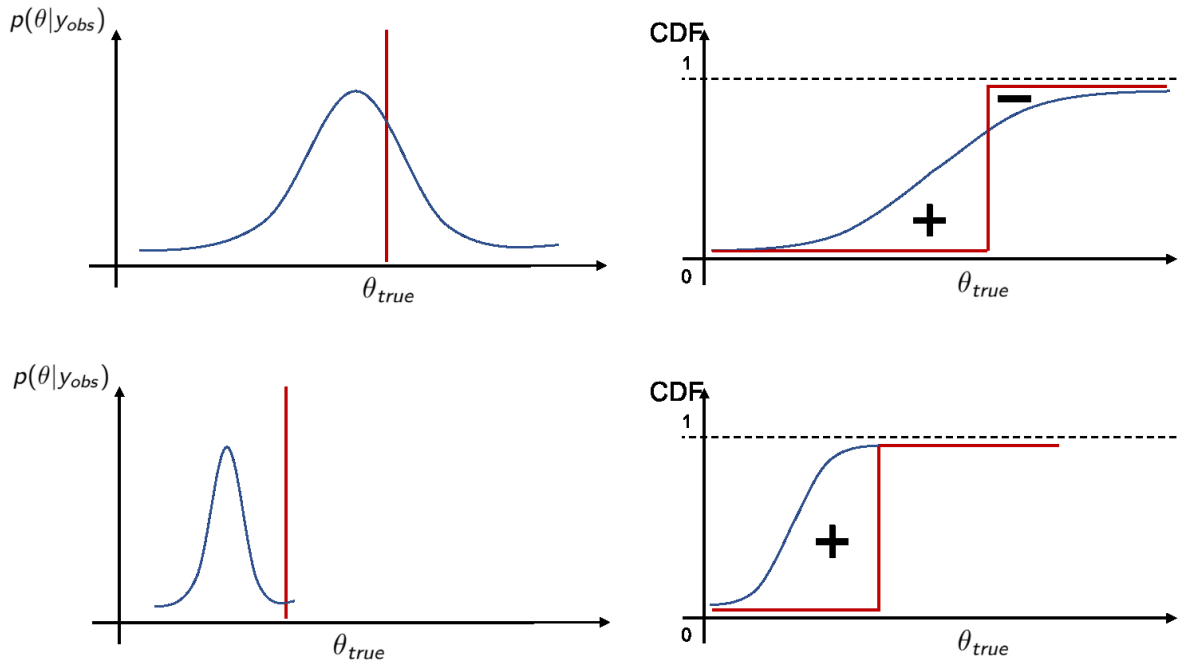}
    \caption{Bias (bottom) and variability (top) quantification in CRPS scores.}
    \label{fig:CRPS_illustr}
\end{figure}

Fairly well known in Bayesian inference are the so called \textit{scoring rules} \cite{gneiting2007strictly}. A score $S(G,H)$ is a measure of predictive accuracy of a forecaster $G$, established through an inference procedure, with respect to $H$, the ``perfect'' forecaster (e.g., true data distribution). A scoring rule is said to be proper if $S(H,H) = \min_G S(G,H) $. In other words, a scoring rule assigns the lowest score to the case where $G$ equals the perfect forecaster. Considering continuous distributions with a density, a perfect forecaster $H$ would be $H(y) = \delta_{y=y_{true}}$, while $G$ can be any distribution $p(y)$ like a posterior distribution.   Some examples of scoring rules are:
\begin{itemize}
    \item \textit{Brier score}(quadratic):
    \begin{equation}
        S(G,H) = \int_{-\infty}^{+\infty} \left( \delta_{y=y_{obs}}(y) - p(y) \right)^2 dy;
    \end{equation}
    \item \textit{Logarithmic score}:
    \begin{equation}
        S(G,H) = - \log p(y_{true});
    \end{equation}
    \item \textit{Continuous ranked probability scores} CRPS :
    \begin{equation}
        S(G,H) = \int_{-\infty}^{+\infty} \left( \int_{-\infty}^y p(z) dz - \mathbbm{1}_{y \leq y_{true}}\right)^2 dy.
    \end{equation}
     Forecasters are CDFs instead of PDFs.
\end{itemize}
All of these rules assign a score zero to the case in which the probability  assigned by $p(y)$ of observing the true data $y_{true}$ is equal to 1. Among these kind of scores of particular interest is for us the case of the CRPS score. This score compares the CDFs of the perfect and inference-built forecasters instead of their PDFs, which presents a number of advantages: since the CDF is a monotone increasing function, subtracting the perfect CDF (a step function set at $y_{true}$) to the inference built CDF provides at the same time a measure of how much bias and variability is contained in the posterior distribution. By bias we mean here how distant is most of the mass of the distribution $p(y)$ from $y_{true}$ and by variability how ``spread-out'' the posterior distribution is. These features are relevant in a data-predictive context in which we want to reproduce data that is as close as possible  to $y_{true}$. Figure \ref{fig:CRPS_illustr} provides a visualization of the concepts behind the CRPS. 

In practice, the real value of $y_{true}$ is unknown and thus the perfect forecaster is approximated by building empirical distributions around ``extra'' or ``newly collected'' data. In the context of our experiment, instead, the CRPS scores would be of direct applicability since we actually know the true value of the quantity of interest $\mathbf{m}_{true}$ (not the data) and the trade-off between bias and variability of the posterior represents a valid way to compare distributions obtained through the two different misfit statistics. 

\subsection{Experimental set-up extension for posterior scoring}

We want to include CRPS scores in our experimental set-up to achieve two objectives:
\begin{enumerate}
    \item Quantitatively assess the performance of the TL-misfit vs. the classic $\ell_2$ distance under different instantiations of $\mathbf{m}_{true}$ for the velocity model configurations $\mathbf{V}_{4_{lay}} \text{ and } \mathbf{V}_{3_{lay}}$ defined in the previous section (experiment \ref{alg:obj-i-CRPS});
    \item Increase the amount of misspecification, by considering $\mathbf{V}_{4_{lay}}$ as the velocity model used for inference, while randomly drawing 20\% variations of $\mathbf{V}_{4_{lay}}$ for the data generating process. In addition to randomizing over the velocity model, we also randomly select a different $\mathbf{m}_{true}$ for each instantiation of misspecified $\mathbf{V}_1$ (experiment \ref{alg:obj-ii-CRPS}).
\end{enumerate}
The second objective is meant to test the robustness of our framework over a larger set of misspecified models. It does not imply considering $\mathbf{V}$ as an additional unknown, jointly with $\mathbf{m}$. Rather, it represents an effort to make sure whatever conclusion we draw for the behavior of the TL misfit is not dependent on the specific realization of $\mathbf{V}_1$ used in experiment 1 or 2. \\
\begin{experiment} 
\SetAlgoLined
 $k=0$\;
\For{$k$ \text{from 1 to}  $  N_{rep} = 1000$}{
 Draw $\mathbf{m}_{true}^k \sim \mathcal{U}(||\mathbf{m}||_{\infty} \leq 1)$\;
 Generate data $\mathbf{y}^k$ \footnote{one vector containing the concatenated signals for each station} according to: \\
\vspace{0.1cm}
\hspace{2cm}
$
\mathbf{y}^k = \mathbf{G}(\mathbf{x}_{true},\mathbf{V}_{4_{lay}},\mathbf{t}) \cdot \mathbf{m}_{true}^{k^{T}} + \mathbf{e} \text{\,\,\, where: } \mathbf{e} \sim \mathcal{N}(0,\sigma^2 \mathbb{I});
$
\vspace{0.2cm}

Estimate the posteriors  $p_{\ell_2}(\mathbf{m}^k \vert \mathbf{y}^k )$ and $p_{\text{TL}_2}(\mathbf{m}^k \vert \mathbf{y}^k )$ assuming the following model for the data:\
\vspace{0.2cm}
\hspace{2cm}
$
\mathbf{u}^k = \mathbf{G}(\mathbf{x}_{true},\mathbf{V}_{3_{lay}},\mathbf{t}) \cdot \mathbf{m}^T + \mathbf{e} \text{\,\,\, where: } \mathbf{e} \sim \mathcal{N}(0,\sigma^2 \mathbb{I}); $
\\
Calculate the CRPS for the $k$-th posterior;\\
}
\caption{Objective (i) - CRPS }
\label{alg:obj-i-CRPS}
\end{experiment}

\begin{experiment} 
\SetAlgoLined
 $k=0$\;
\For{$k$ \text{from 1 to}  $  N_{rep} = 1000$}{
 Draw $\mathbf{m}_{true}^k \sim \mathcal{U}(||\mathbf{m}||_{\infty} \leq 1)$\;
 Draw $\mathbf{V}_{4-20\%}^k \sim \mathcal{U}(\{\mathbf{V} \,\,\, s.t. \,\,\, 0.8V_{1_{p}} \leq V_p \leq 1.2V_{1_{p}} ,  0.8V_{1_{s}} \leq V_s \leq 1.2V_{1_{s}}  \,\, \forall \, \, V_p, V_s \in \mathbf{V} \text{ and } V_{1_{p}}, V_{1_{s}} \in \mathbf{V}_1 \})$\;
 Generate data $\mathbf{y}^k$ \footnote{one vector containing the concatenated signals for each station} according to: \\
\vspace{0.1cm}
\hspace{2cm}
$
\mathbf{y}^k = \mathbf{G}(\mathbf{x}_{true},\mathbf{V}_{4_{lay}},\mathbf{t}) \cdot \mathbf{m}_{true}^{k^{T}} + \mathbf{e} \text{\,\,\, where: } \mathbf{e} \sim \mathcal{N}(0,\sigma^2 \mathbb{I});
$
\vspace{0.2cm}

Estimate the posteriors $p_{\ell_2}(\mathbf{m}^k \vert \mathbf{y}^k )$ and $p_{\text{TL}_2}(\mathbf{m}^k \vert \mathbf{y}^k )$ assuming the following model for the data:\
\vspace{0.2cm}
\hspace{2cm}
$
\mathbf{u}^k = \mathbf{G}(\mathbf{x}_{true},\mathbf{V}_{4-20\%}^k,\mathbf{t}) \cdot \mathbf{m}^T + \mathbf{e} \text{\,\,\, where: } \mathbf{e} \sim \mathcal{N}(0,\sigma^2 \mathbb{I}); $
\\
Calculate the CRPS for the $k$-th posterior;\\
}
\caption{Objective (ii) - CRPS }
\label{alg:obj-ii-CRPS}
\end{experiment}

\subsubsection{Objective (i)} 
\begin{figure}
    \centering
    \includegraphics[width=\textwidth]{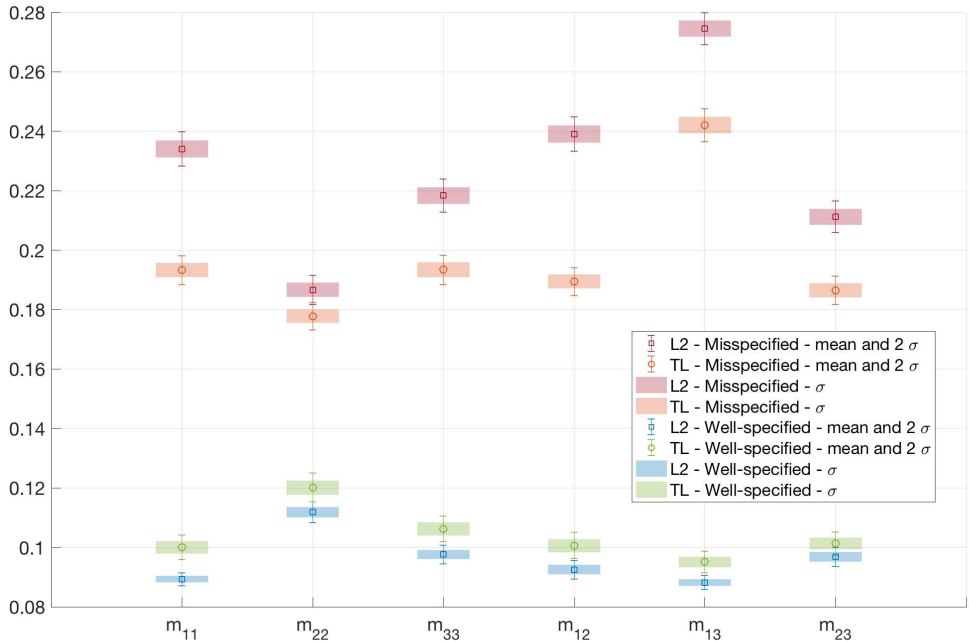}
    \caption{Mean CRPS scores in the well-specified (WS) and misspecified (MS) case and relative error bars.}
    \label{fig:Mean_CRPS_MISP_ErrBar}
\end{figure}

\begin{figure}
    \centering
    \includegraphics[width=\textwidth]{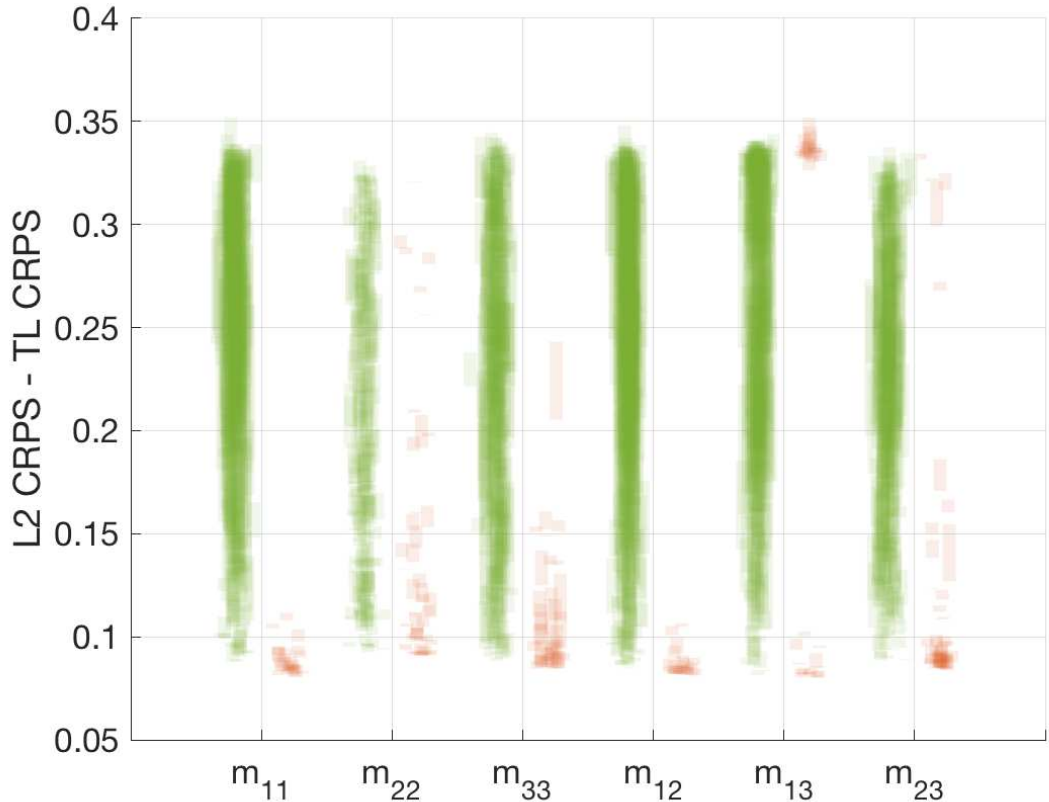}
    \caption{Box-plot for $\Delta_k$ for each moment tensor component in experiment \ref{alg:obj-i-CRPS}.}
    \label{fig:Experiment4_Error_bars}
\end{figure}

The results from this experiment can be analyzed in multiple ways, each revealing different pieces of information. First, for each of the posteriors obtained in experiment \ref{alg:obj-i-CRPS} we can calculate the CRPS score as follows:
\begin{equation}
    \text{CRPS} = \frac{1}{N} \sum_{i}^N \left( F(\mathbf{m}_i \vert \mathbf{y}_{obs}) - \mathbbm{1}_{\mathbf{m}_i > \mathbf{m}_{true}} \right)^2,
\end{equation}
where $F$ is the empirical cumulative distribution function of a given posterior and the step function is the ideal CDF for the true value of moment tensor. As a first comparison measure we calculate the mean CRPS for each of the moment tensor components obtained  for both the $\ell_2$ and TL-based posteriors. 
\begin{equation}
    \overline{\text{CRPS}} = \frac{1}{N_{rep}} \sum_{k=1}^{N_{rep}} \text{CRPS}_k.
\end{equation}
We report them in Figure \ref{fig:Mean_CRPS_MISP_ErrBar} together with the associated estimator variance:
\begin{equation}
    \sigma_{\overline{\text{CRPS}}}= \frac{1}{\sqrt{N_{rep}}} \sqrt{ \sum_{k=1}^{N_{rep}} \frac{ (\text{CRPS}_k - \overline{\text{CRPS}})^2 }{N_{rep}-1}}.
\end{equation}
In order to make the comparison more significant we have repeated  experiment \ref{alg:obj-i-CRPS} in a well-specified setting i.e. with both the data and inference model Green's functions set to $ \mathbf{G}(\mathbf{x}_{true},\mathbf{V}_{4_{lay}},\mathbf{t})$ and while using both the $\ell_2$ and TL distance as misfit statistics. \\
{}
While in the well-specified setting both distances exhibit similar low scores, in the misspecified settings the difference between the scores obtained with the $\ell_2$ distance and those obtained with the TL distance is quite significant. The distributions $p_{\ell_2}^1$ and $p_{\ell_2}^2$ achieve on average higher scores than the $p_{TL}^1$ and $p_{TL}^2$, which indicates higher bias and/or variability. \\
Average scores however do not provide a comprehensive image of the TL vs. $\ell_2$ performance in terms of uncertainty quantification. Assuming that, as shown by the mean values, their behavior is almost identical in the well specified case, we focus on the misspecified setting. In this case,  we are particularly interested in  answering the following question: given the same $\mathbf{m}_{true}$ and the same velocity misspecification, how do the CRPS associated to the $\ell_2$-based posterior compare to those obtained in the $\text{TL}_2$-based one? In particular, are the CRPS scores obtained for the $p_{\ell_2}(\mathbf{m}^k \vert \mathbf{y}^k )$ always higher than those obtained for the $p_{\text{TL}_2}(\mathbf{m}^k \vert \mathbf{y}^k )$? To provide a comprehensive and visual answer to this question we build the graph in Figure \ref{fig:Experiment4_Error_bars}. For some randomly sampled pairs of CRPS, and every component of the moment tensor, we calculate the relative difference $\Delta_k$ and mid-point $\overline{\Delta}_k$:
\begin{align}
    \Delta_k &= \text{CRPS}_k^{\ell_2} - \text{CRPS}_k^{\text{TL}_2} \\
    \overline{\Delta}_k &= \frac{\text{CRPS}_k^{\ell_2} + \text{CRPS}_k^{\text{TL}_2} }{2}
\end{align}
We then graph this information in the following fashion:
\begin{enumerate}
    \item we select the moment tensor component of interest (horizontal axis);
    \item if $\Delta_k \geq 0$, we plot a green box of height $\Delta_k$ with the centroid y-coordinate set at $\overline{\Delta}_k$. The width of the box is set to fixed value for graphical purposes only;
    \item if $\Delta_k < 0$, we proceed as above, except that we will use the color red;
    \item each box is filled with translucent color, which will produce darker shaded regions where multiple $\Delta_k$ will be centered around.
\end{enumerate}
The characteristics of this plot allow for the following interpretation: the boxes being translucent, if in the majority of cases the difference in scores between the $\ell_2$ and $\text{TL}_2$ posteriors is positive, then we will see a darker shade of green above the $x$-axis. If instead  the difference in scores is predominantly negative, the we will see darker shades of red. Positive $\Delta_k$ are predominant for all moment tensor components, showing a superior performance in terms of predictive capability of the TL-based posteriors. 

{}

\begin{figure}
    \centering
    \includegraphics[width=\textwidth]{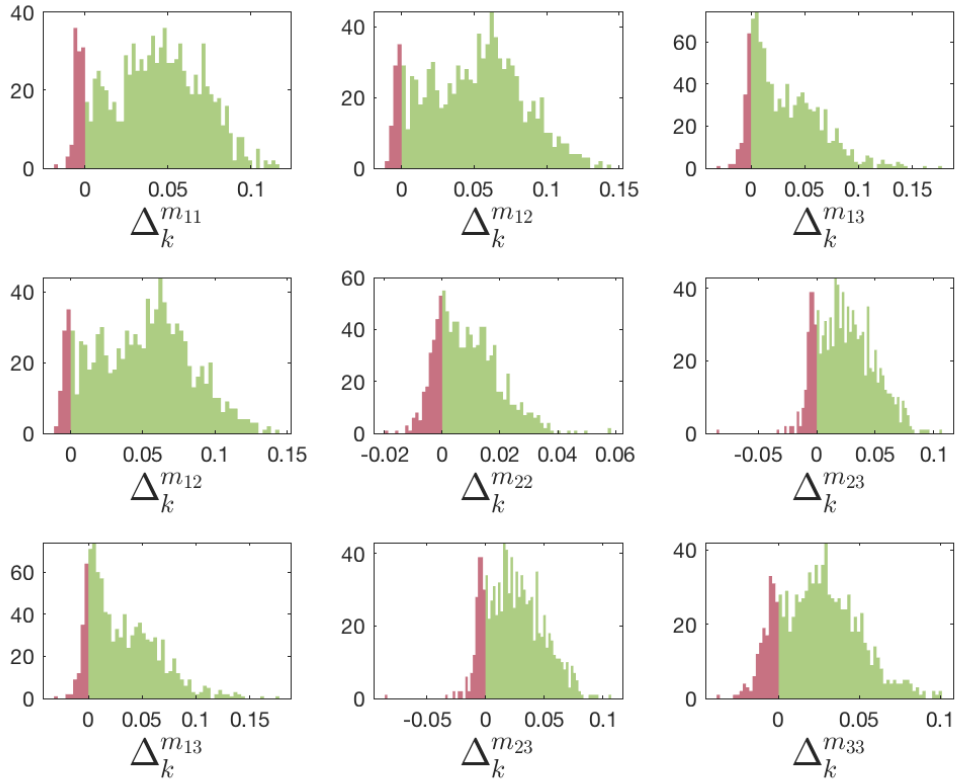}
    \caption{Histograms for $\Delta_k$ for each moment tensor component, arranged in a moment tensor matrix format for experiment \ref{alg:obj-i-CRPS}.}
    \label{fig:Experiment4_M_tensor_matrix}
\end{figure}

\begin{figure}
    \centering
    \includegraphics[width=\textwidth]{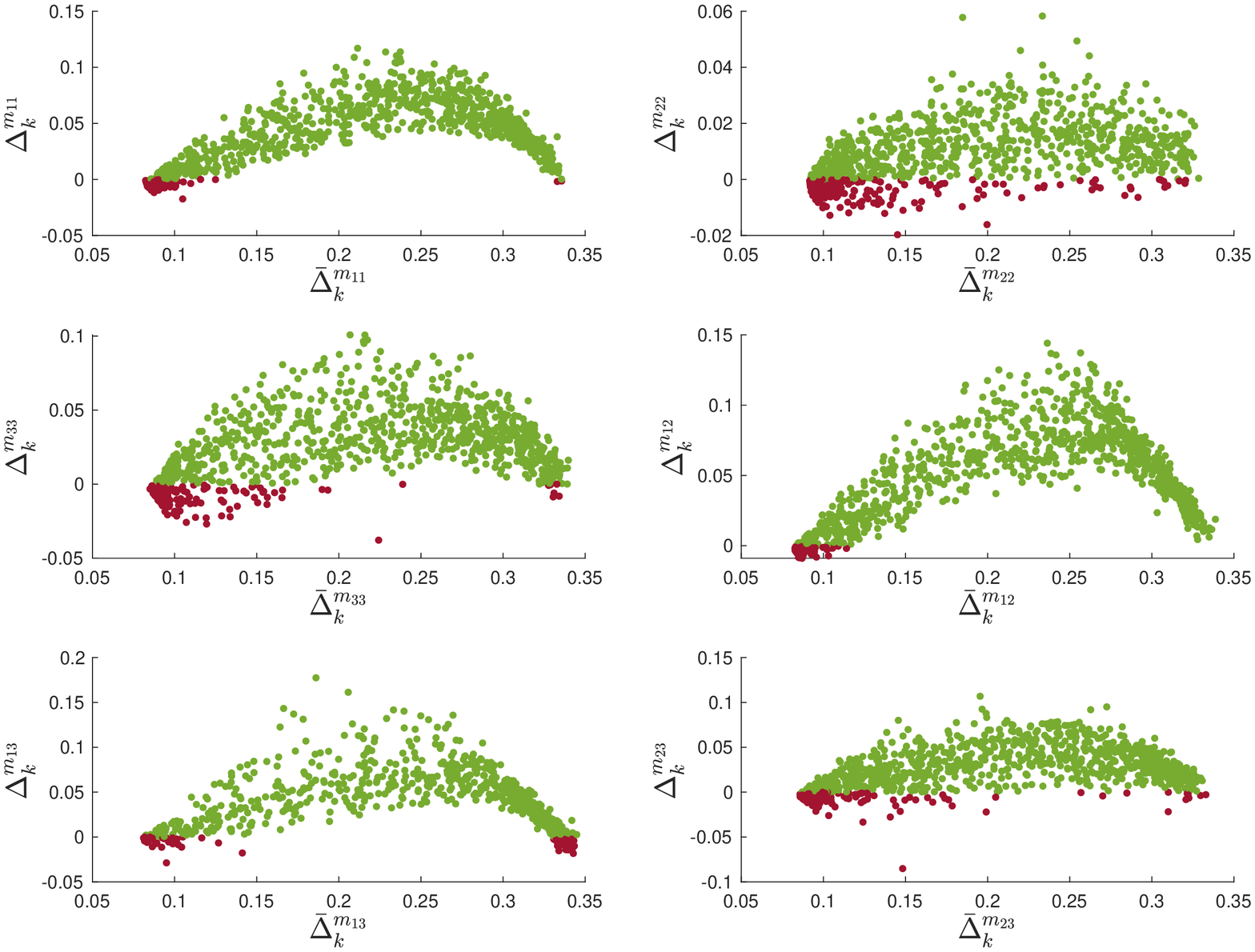}
    \caption{Scatter plot of $\Delta_k$ v.s. y-coordinate set at $\overline{\Delta}_k$ for experiment \ref{alg:obj-i-CRPS}.}
    \label{fig:Experiment4_Scatter_plot}
\end{figure}

To achieve an even deeper analysis of the TL vs. $\ell_2$ performance as a misfit, we also plotted a histogram of the $\Delta_k$ per each moment tensor component (color-coded in the histogram) as well as a scatter plot of the $\Delta_k$ v.s. $\bar{\Delta}_k$. While the histograms in Figure \ref{fig:Experiment4_M_tensor_matrix} clearly confirm the prevailing positive nature of the $\Delta_k$ already discussed in Figure \ref{fig:Experiment4_Error_bars}, the scatter plots in Figure \ref{fig:Experiment4_Scatter_plot} provide additional information on the distribution of the $\Delta_k$ and the respective average score values $\bar{\Delta}_k$. We conclude by tabulating the value of the estimated mean of $\Delta_k$ and relative standard deviation for each moment tensor component (Table \ref{tab:Fixed_V_Mean_Diff}).

{}

{}

\begin{table}[H]
\centering
\begin{tabular}{||c c c c c c c||}
\hline
Parameter & $m_{11}$ & $m_{22}$ & $m_{33}$ & $m_{12}$  & $m_{13}$ & $m_{23}$  \\ [0.5ex] 
\hline\hline
\textbf{Mean $\Delta_k$} &  0.0407 &   0.0089  &  0.0249 &   0.0496  &  0.0325  &  0.0247 \\
\hline
\textbf{Std.  } &0.0009  &  0.0003  &  0.0007  &  0.0010  &  0.0010  &  0.0007
 \\ 
\hline
\end{tabular}
\caption{Mean $\Delta_k$ values and associated estimator standard deviation - experiment \ref{alg:obj-i-CRPS}}
\label{tab:Fixed_V_Mean_Diff}
\end{table}

\subsubsection{Objective (ii)} 
\begin{figure}
    \centering
    \includegraphics[width=\textwidth]{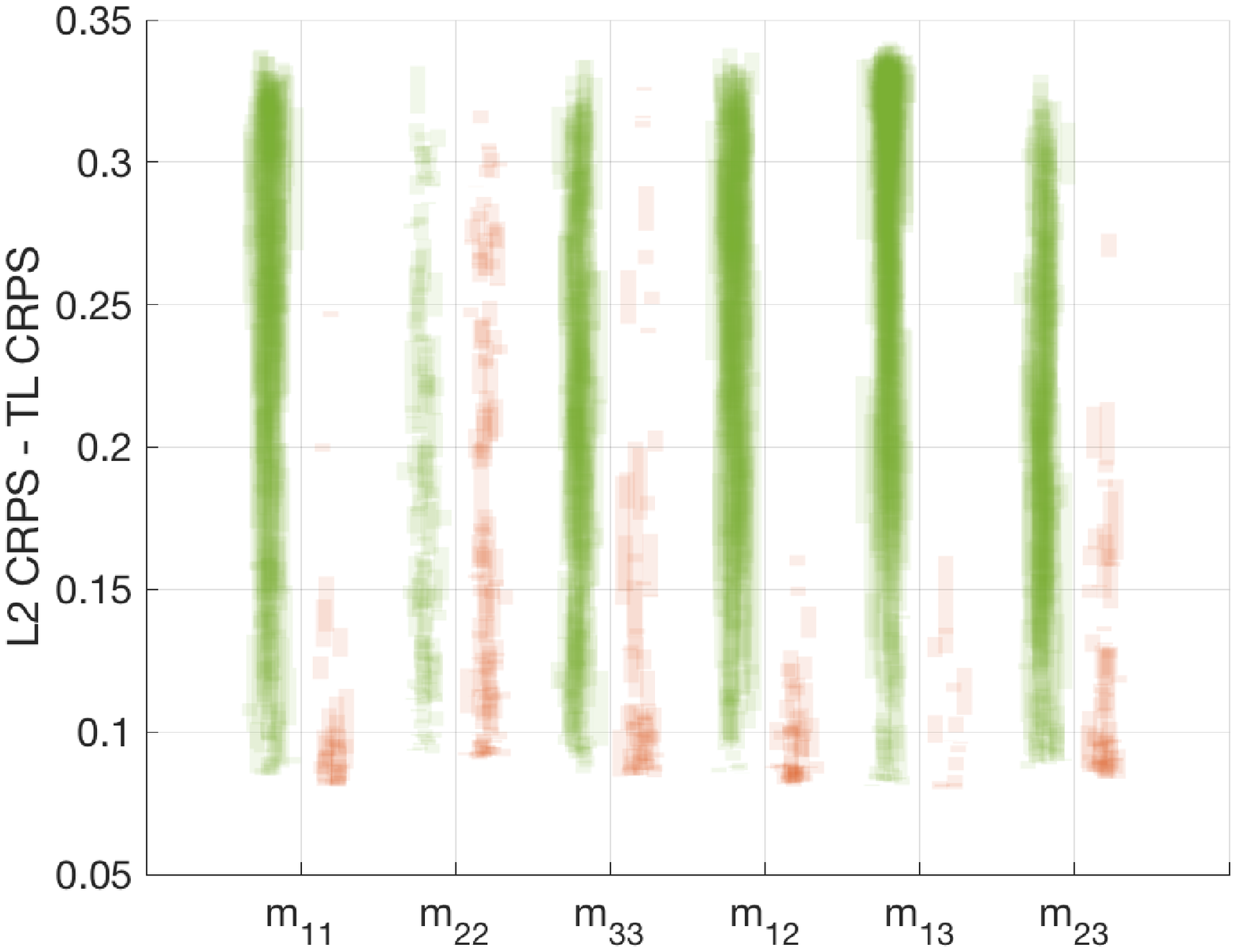}
    \caption{Box-plot for some randomly sampled $\Delta_k$ for each moment tensor component in experiment \ref{alg:obj-ii-CRPS}. Green boxes correspond to positive differences, red boxes to the negative ones.}
    \label{fig:Error_Bar_Rand_V}
\end{figure}

\begin{figure}
    \centering
    \includegraphics[width=\textwidth]{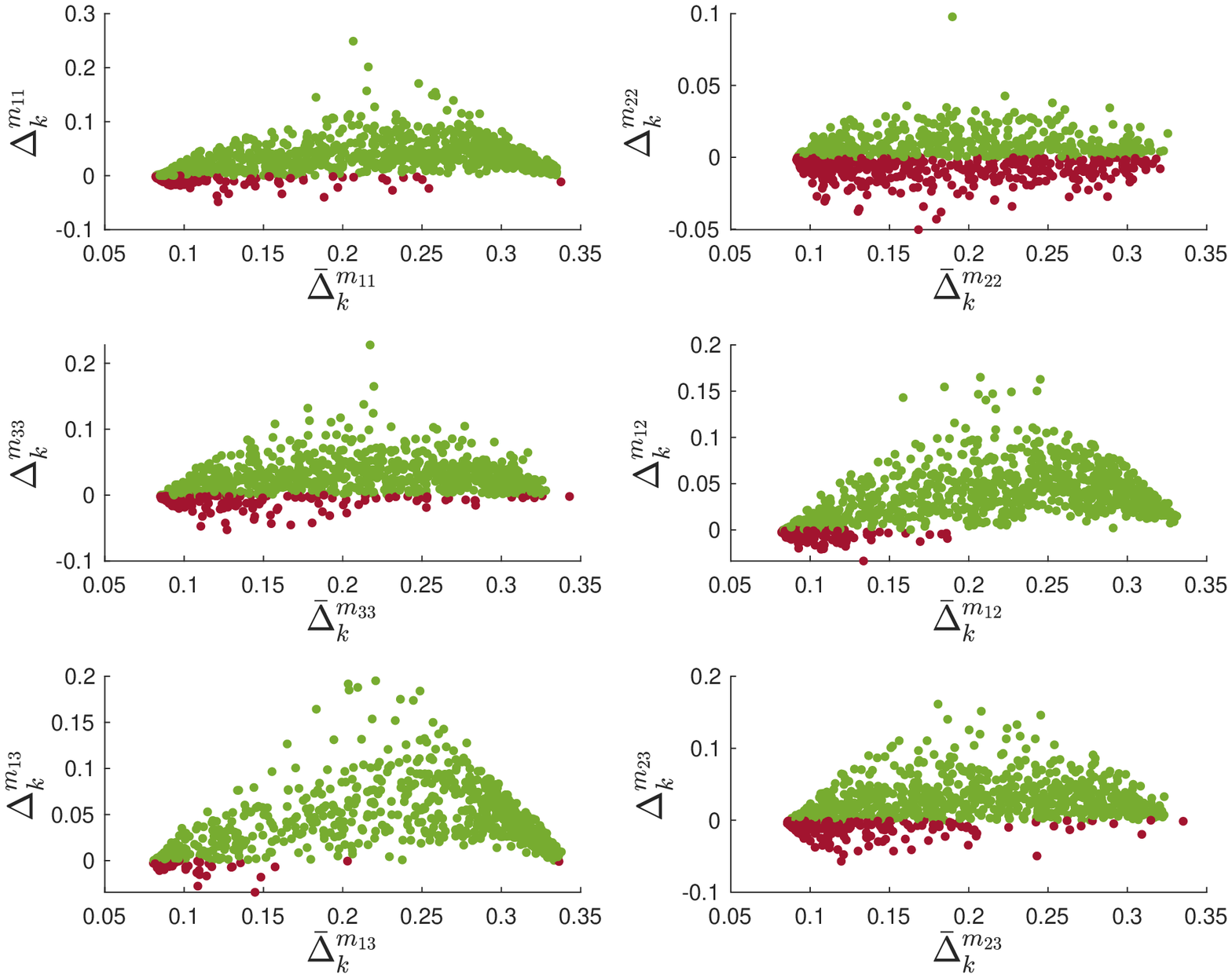}
    \caption{Scatter plot of $\Delta_k$ v.s. y-coordinate set at $\overline{\Delta}_k$ for experiment \ref{alg:obj-ii-CRPS}. }
    \label{fig:scatter_Rand_V}
\end{figure}

\begin{figure}
    \centering
    \includegraphics[width=\textwidth]{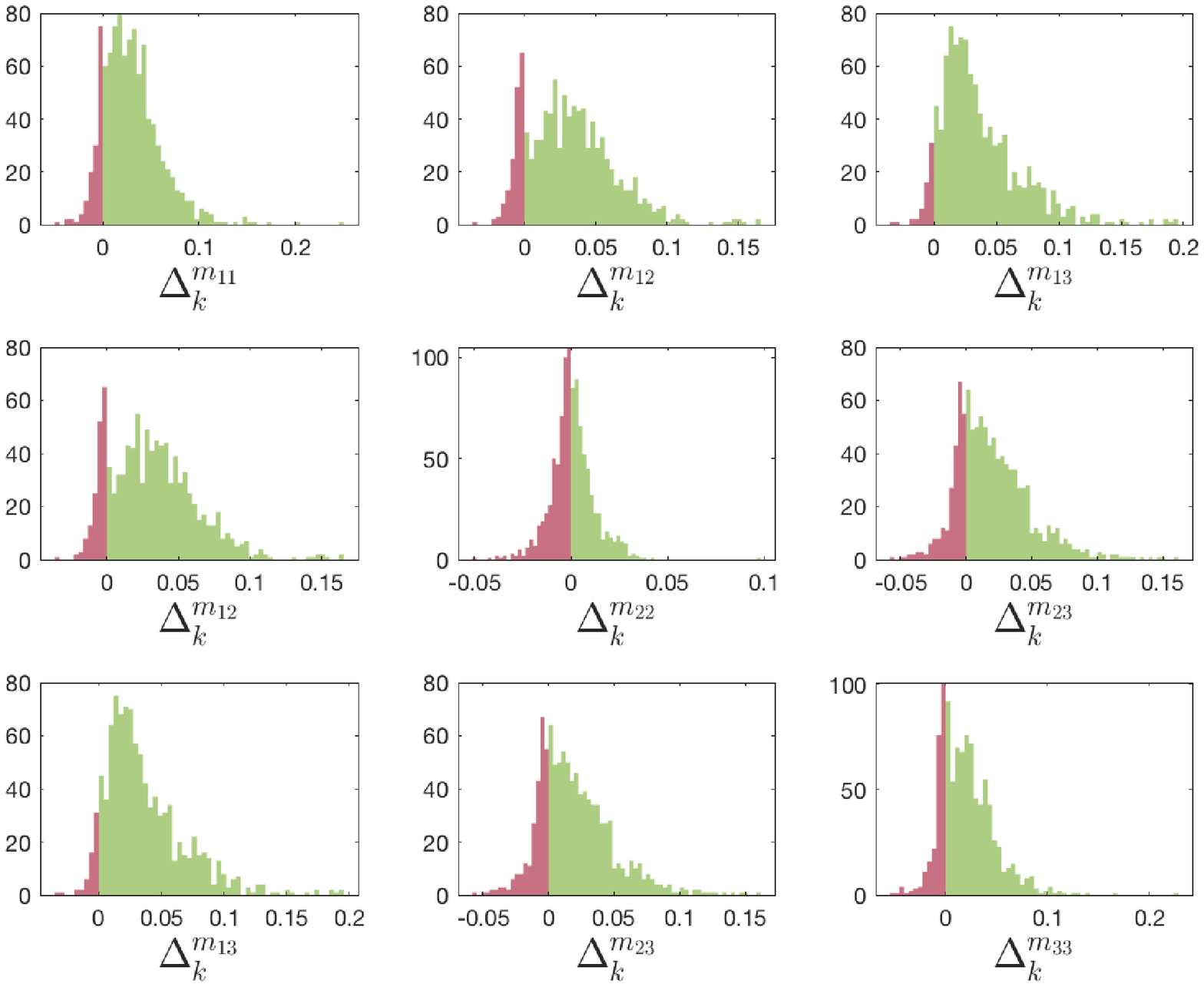}
    \caption{Histograms for $\Delta_k$ for each moment tensor component, arranged in a moment tensor matrix format for experiment \ref{alg:obj-ii-CRPS}.}
    \label{fig:Hist_Rand_V}
\end{figure}

In this second set of experiments we broadened the spectrum of the misspecification over the velocity model as described in detail in experiment \ref{alg:obj-ii-CRPS}. To analyze the results in this setting we will only focus on the plots characterizing  the differences in CRPS obtained from $\ell_2$ and $\text{TL}_2$-based posteriors. Figure \ref{fig:Error_Bar_Rand_V} shows a general prevalence of positive $\Delta_k$ for almost all moment tensor components, except $m_{22}$, where the distribution of the $\Delta_k$ is almost symmetric around the $x$-axis. This is also reflected in the scatter plot in Figure \ref{fig:scatter_Rand_V} as well as the histograms for each moment tensor component (Figure \ref{fig:Hist_Rand_V}). These observations are also confirmed by the mean $\Delta_k$ scores and associated estimator variances (table \ref{tab:Rand_V_Mean_Diff}). The means for all components except $m_{22}$ are positive and the estimator variance sufficiently small to guarantee it. 

{}

{}

{}

\begin{table}[H]
\centering
\begin{tabular}{||c c c c c c c||}
\hline
Parameter & $m_{11}$ & $m_{22}$ & $m_{33}$ & $m_{12}$  & $m_{13}$ & $m_{23}$  \\ [0.5ex] 
\hline\hline
\textbf{Mean $\Delta_k$} &0.0297 &  0.0007  & 0.0201  &  0.0319 &  0.0366 &  0.0200\\
\hline
\textbf{Std.  } & 0.0010  &  0.0004  &  0.0009  &  0.0009  &  0.0010  &  0.0009 \\ 
\hline
\end{tabular}
\caption{Mean $\Delta_k$ values and associated estimator standard deviation - experiment \ref{alg:obj-ii-CRPS}}
\label{tab:Rand_V_Mean_Diff}
\end{table}

\subsubsection{Conclusions}It appears that the newly introduced misfit function is able to minimize the impact of velocity misspecification while inferring the moment tensor components. As shown in previous literature \cite{yang2018application} this distance is fairly insensitive to translation of the signals, which is the most likely impact a misspecified velocity may have on a waveform. In a broader perspective, this confirms that a non-sufficient misfit statistic can be used to disregard parts of the data-set that are most affected by model misspecification and are not particularly relevant to estimate the quantity of interest. \\

\subsection{Hierarchical model and analytical solution} \label{sec:analytical}
\begin{figure}
    \centering
    \includegraphics[width=\textwidth]{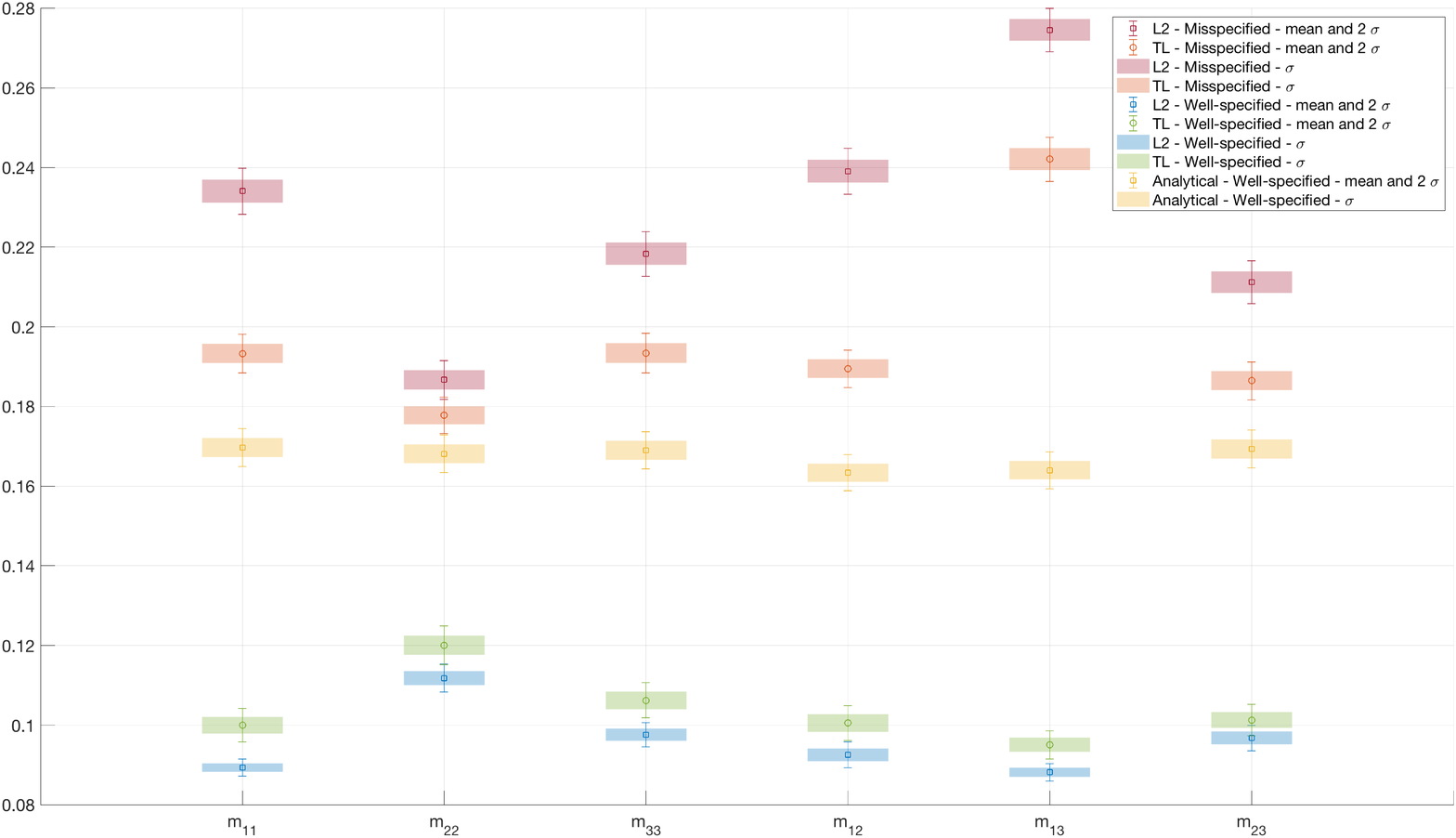}
    \caption{Mean CRPS scores in the well-specified (WS) and misspecified (MS) case and relative error bars, with the additional analytic solution to the WS case.}
    \label{fig:Mean_CRPS_MISP_ErrBar_plus_analy}
\end{figure}

\begin{figure}
    \centering
    \includegraphics[width=0.8\textwidth]{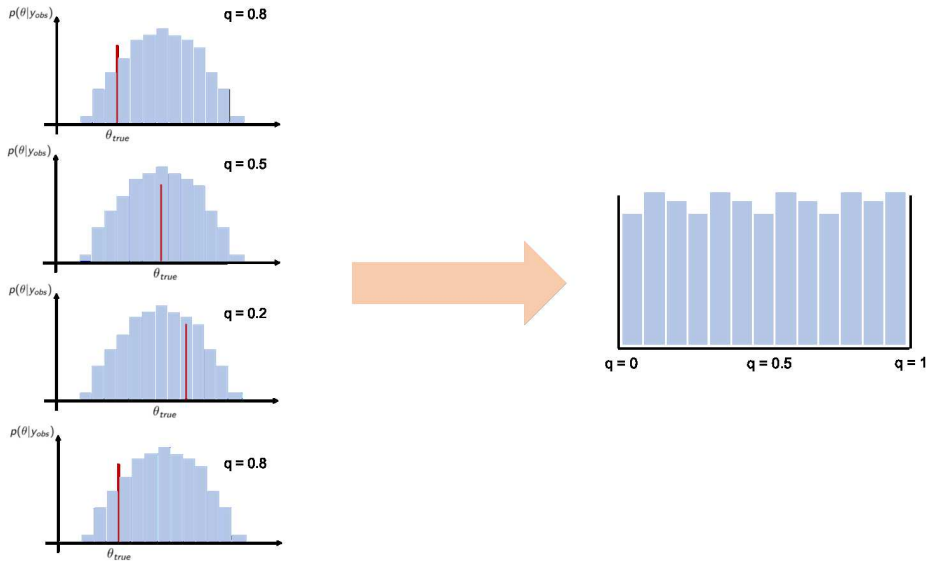}
    \caption{Quantile rank histogram building process under consistent conditions.}
    \label{fig:rank_hist}
\end{figure}

\begin{figure}
\centering
\begin{subfigure}[b]{.4\textwidth}
        \centering
        \includegraphics[width=5cm]{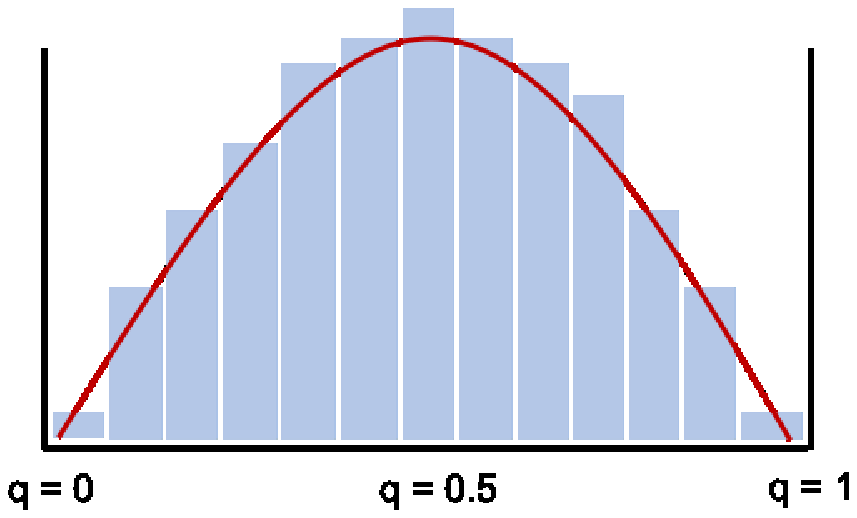}
        \caption{Bell-shaped  indicates an over-dispersed forecasting distribution.}
        \label{fig:QRANK_OVER}
\end{subfigure}%
~
\begin{subfigure}[b]{.4\textwidth}
    \centering
    \includegraphics[width=5cm]{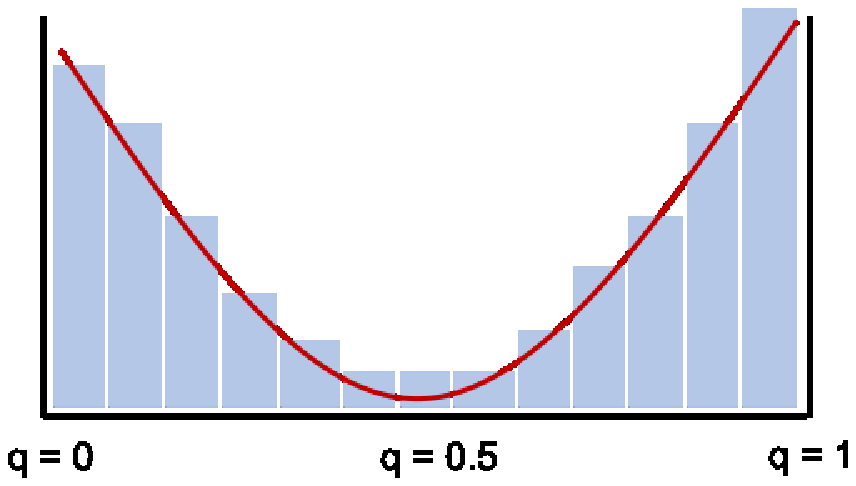}
    \caption{U-shaped indicates a biased forecasting distribution.}
    \label{fig:QRANK_UNDER}
\end{subfigure}
\caption{Non-uniform quantile rank-histogram shapes.}
\end{figure}

\begin{figure}
    \centering
    \includegraphics[width=\textwidth]{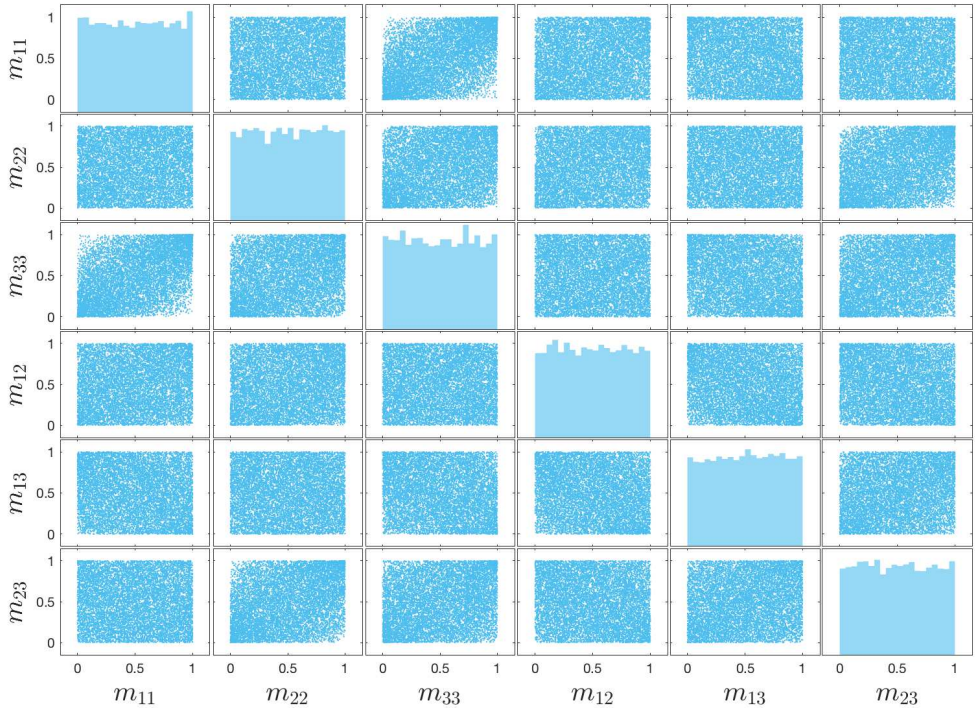}
    \caption{Quantile rank-histogram analytical model - well specified setting.}
    \label{fig:q_rank_analy}
\end{figure}

\begin{figure}
    \centering
    \includegraphics[width=\textwidth]{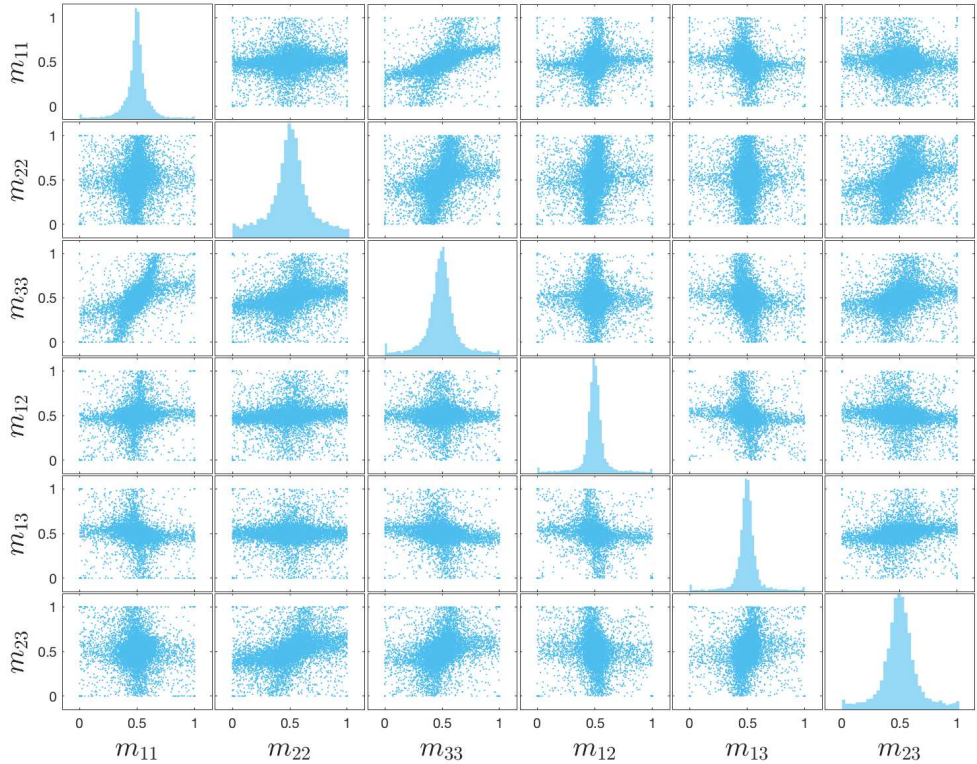}
    \caption{Quantile rank-histogram hierarchical model - well specified setting.}
    \label{fig:q_rank_hier}
\end{figure}

\begin{figure}
    \centering
    \includegraphics[width=\textwidth]{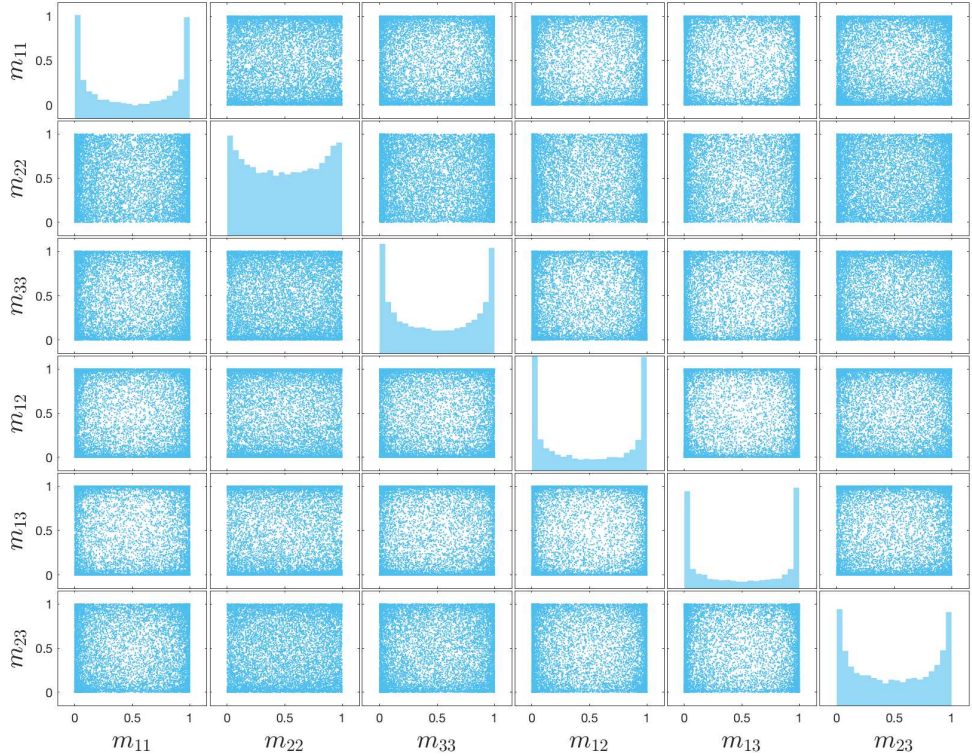}
    \caption{Quantile rank-histogram hierarchical model - misspecified setting $\ell_2$. }
    \label{fig:L2_MCMC_10000}
\end{figure}

\begin{figure}
    \centering
    \includegraphics[width=\textwidth]{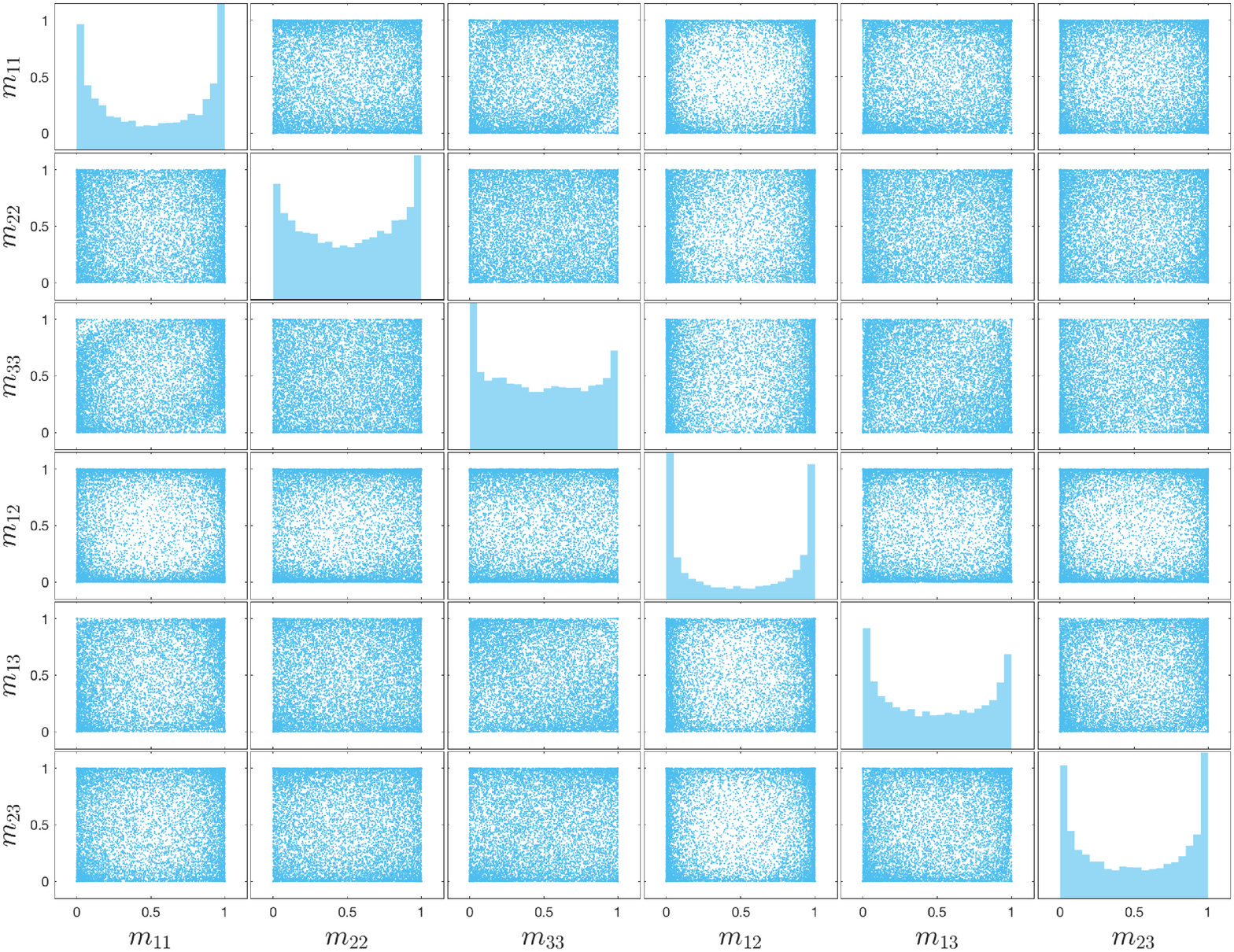}
    \caption{Quantile rank-histogram hierarchical model - misspecified setting $\text{TL}_2$ .}
    \label{fig:TL_10000_MCMC_2}
\end{figure}

In the previous section, we mentioned that in a well-specified setting it is possible to obtain an analytical solution to the linear-Gaussian inverse problem.  In fact, assuming the noise level is known and fixed (i.e. $ \Sigma = \sigma^2  \mathbb{I} $) and an unbounded improper uniform prior, the posterior distribution is a truncated Gaussian with mean and variance as follows:
\begin{equation}
    \mathbf{m} \vert \mathbf{y}, G(\mathbf{x_{true},V_{true}}) \sim \mathcal{N}((G^T G)^{-1}G^T \mathbf{y}, \sigma^2(G^T G)^{-1}).
\end{equation}
Intuitively, this model should exhibit less variability with  respect to the corresponding hierarchical one since the variance is known and there is also no need to add the scaling parameter $s$. The likelihood function inherently handles the scaling of the data-model misfit. We therefore repeat experiment \ref{alg:obj-i-CRPS} with the well specified velocity model and by using the analytic solution for the inverse problem. We then proceed with the calculation of the mean CRPS scores for this newly obtained set of posterior distributions and plot them against the ones coming from the hierarchical models (Figure \ref{fig:Mean_CRPS_MISP_ErrBar_plus_analy}). While the analytical solution scores are expectedly lower than the ones obtained in the misspecified case, they appear significantly higher than those obtained in the well-specified case with a hierarchical model. \\
{}
This result is somewhat counter-intuitive since it seems to convey the message that a more uncertain model produces better posteriors than a model that, from a theoretical standpoint, should be more thorough. This apparent contradiction can instead be explained by referring to a discussion which is central to this paper: there never is a ``perfect'' posterior, but rather a good posterior for a determined objective. Scoring rules tend to reward the predictive capability of a posterior, which is achieved with the least amount of bias and  variability as possible. However, these properties may not completely overlap with  other, statistically consistent, behaviors of posterior distributions. In particular a perfect forecaster may not exhibit what is known as the frequentist behavior of Bayesian credible intervals. In order to describe what this behavior is, we introduce another type of posterior-check: posterior quantile rank statistics \cite{talts2018validating} \cite{cook2006validation}. We stress that, contrary to the CRPS score, this is a \textit{self-consistency} test that aims at answering the following question: is there any inherent bias in the way the posterior characterizes the uncertainty around the parameter space? Experiment \ref{alg:rank_stat} describes the steps necessary to calculate quantile rank statistics and associated histograms for our specific test-case.
The last step of this algorithm aims at verifying that the true values of $\mathbf{m}$ fall uniformly across the posterior credible set, just as, in a frequentist setting, an $\alpha-$confidence interval contains the true value $\alpha-\%$ of the times (Figure \ref{fig:rank_hist}). This behavior translates into a uniform histogram over the sampled values of $q$: it indicates  that the posterior distributions are neither overly biased towards one subset of the parameter set (Figure \ref{fig:QRANK_UNDER}), nor  overly dispersive (Figure \ref{fig:QRANK_OVER}), over-representing the amount of uncertainty in the problem. \\
\begin{experiment}[h] 
\SetAlgoLined
 $k=0$\;
\For{$k$ \text{from 1 to}  $  N_{rep} = 1000$}{
 Draw $\mathbf{m}_{true}^k \sim \mathcal{U}(||\mathbf{m}||_{\infty} \leq 1)$\;
 Generate data $\mathbf{y}^k$  according to: \\
\vspace{0.1cm}
\hspace{2cm}
$
\mathbf{y}^k = \mathbf{G}(\mathbf{x}_{true},\mathbf{V}_{true},\mathbf{t}) \cdot \mathbf{m}_{true}^T + \mathbf{e} \text{\,\,\, where: } \mathbf{e} \sim \mathcal{N}(0,\sigma^2 \mathbb{I});
$
\vspace{0.2cm}

Estimate the posterior $p(\mathbf{m}^k \vert \mathbf{y}^k )$ assuming the following model for the data $\mathbf{u}^k = \mathbf{G}(\mathbf{x}_{true},\mathbf{V}_{true},\mathbf{t}) \cdot \mathbf{m}^T + \mathbf{e} \text{\,\,\, where: } \mathbf{e} \sim \mathcal{N}(0,\sigma^2 \mathbb{I}) $\;
Draw M samples $\mathbf{m}_i$ from the posterior distribution $p(\mathbf{m}^k \vert \mathbf{y}^k )$\;
Calculate: $q_k = \frac{1}{M} \sum_{i=1}^M \mathbbm{1}_{\theta_i > \theta_{true}} $\;
}
Verify that $q \sim \mathcal{U}[0,1]$; 
\caption{Quantile rank statistics}
\label{alg:rank_stat}
\end{experiment}
We plotted  the quantile rank statistics histogram for our experiment in the well specified case when using the $\ell_2$ misfit both with the hierarchical model (Figure \ref{fig:q_rank_hier}) and with the  analytical  solution (Figure \ref{fig:q_rank_analy}). It can be observed that while with the analytical solution the histogram is uniform as expected, in the hierarchical model case it assumes a relatively narrow delta-shape around the center value $0.5$. This result offers an explanation for the apparent inconsistency between the associated CRPSs: the analytical solution scores higher than the hierarchical model since it is behaves ``perfectly'' in Bayesian terms i.e. exhibits good frequentist coverage over repeated realizations. Concretely, this means that the true value does not always sit in the middle of the posterior distribution, which induces a significant penalty in terms of bias. The hierarchical model posterior instead is more consistently centered around  $\mathbf{m}_{true}$, lowering the CRPS score at the price of over-dispersion. \\
For completeness it is also interesting to check what the quantile rank histograms look like in the misspecified case as well. Figures \ref{fig:L2_MCMC_10000} and \ref{fig:TL_10000_MCMC_2}  show the histograms in these cases. The results can be interpreted in the following way: under conditions of model misspecification the posterior distributions are, on average, biased and concentrate around the wrong values often enough for the histogram to assume the characteristic U-shape. This behavior is consistent with the Bernstein-Von Mises theorem discussed in section \ref{sec:LitRewModelMis}. While the histograms under these two cases look fairly similar it is worth nothing that the almost uniform histogram for component $m_{22}$ in the $\ell_2$ case is a byproduct of the fact that the associated posteriors are almost uniform. In fact, when a posterior is always uniform (bounded) and the true value is drawn from a uniform (bounded) prior, then the relative quantile rank histogram will also always be uniform. This may once again appear as a contradiction between quantile-rank checks that reward a totally uninformative posterior versus another kind of posterior checks (CRPS) that instead penalize the same posterior, since it is unusable from a forecasting point of view. \\
Regardless of the specific moment tensor component, the CRPS clearly highlight a  difference between the quality of the posterior distributions obtained with the $\ell_2$ distance and the ones based on the $\text{TL}_2$ in a misspecified setting. However, the quantile rank histograms only slightly favor the use of optimal transport. This indicates that while the $\text{TL}_2$ can make inference more robust to misspecification in terms of predictive capabilities, it does not eliminate the misspecification itself.\\
Trough this last exercise we have observed that that different posterior quality checks reward different behaviors of the posterior and that even a ``perfect'' posterior from a theoretical standpoint, may be negatively rewarded by a scoring system that only focuses on its predictive capability. In this regard, the rank-histograms are a more comprehensive measure of the  ``correctness'' of an inference framework compared to the CRPS scores. However, as it often occurs in science and engineering, a ``wrong'' model may be more useful, under specific circumstances, than a theoretically ``sound'' one.

\section{Conclusions and perspectives}
This paper discusses the potential benefit of using transport-Lagrangian distances under model misspecification in a Bayesian framework. This alternative distance, as opposed to the more classic $\ell_p$ norms, is tested in the context of moment tensor inversion, a seismic inverse problem. The TL-misfit appears to be robust to translations of  signals due to a misspecified velocity model for the propagation of seismic waves. This induces, on average, better posteriors in terms of predictive capability than those obtained by building a likelihood function directly off an additive Gaussian noise assumption. This choice leads to looking at the signed differences between model predicted and observed waveforms while implicitly inducing an $\ell_2$-misfit. The term $\mathbf{y-u}$ is a sufficient statistic i.e. no amount of information coming from the data is disregarded when inferring the quantity of interest. The TL distance, instead,  acts as a purposely non-sufficient statistic that disregards non-relevant information to the inference of moment tensor components that also happen to be the data-features most affected by the misspecification of the velocity model. 
This paper also presents a coherent framework to adopt the TL distance as a misfit measure in a Bayesian setting.

Although the findings  are limited to a relatively simple application,  we believe alternative likelihood functions like the TL-distance constitute the foundations for a more general approach to Bayesian inverse problems in the presence of model misspecification. Retaining or dismissing only a portion of the information contained in the model-data discrepancy by  carefully choosing a non-sufficient misfit statistic can be beneficial to a broader class of problems that exhibit the same structure of the moment tensor inversion one: the quantity of interest is easily associable with a particular aspect of the data that is well represented in the model space, while some other data-features cannot be captured by the model and therefore only represent a nuisance factor when performing inference. The claim is that alternative misfit measures can be tailored to bring the data inside the model space, effectively nullifying the effect of misspecification, while keeping instead the relevant information to estimate the quantity of interest.\todo{A lot of this paragraph would be better off in the conclusions section!}
The TL-distance seems promising as to how to flexible it can be in terms of rewarding  certain discrepancies between signal data-type. A flexibility well represented not only by the customizable parameter $\lambda$ (that controls how much transport of the signal should occur across the time domain), but also the choice of the cost function itself. Additional work is also undergoing to identify proper TL-based likelihood functions.


\clearpage
\noindent\textbf{References}\\
\bibliographystyle{plain}
\bibliography{biblio} 

\end{document}